\documentclass[journal]{IEEEtran}
\ifCLASSINFOpdf
\else
\fi
\usepackage{amsmath,amssymb,amsfonts}
\usepackage{algorithmic}
\usepackage{graphicx}
\usepackage{epstopdf}
\usepackage{textcomp}
\usepackage{xcolor}
\usepackage{mathtools}
\usepackage[ruled,linesnumbered,noend]{algorithm2e}
\usepackage{caption}
\usepackage{tabularray}
\usepackage{subcaption}
\usepackage{booktabs}
\usepackage{enumerate}
\usepackage{rotating}
\usepackage{bbding}
\usepackage{pifont}
\usepackage{wasysym}
\usepackage{xspace}

\usepackage[nolist,nohyperlinks]{acronym}
\newcommand{\po}[2]{P^{\mathcal{#1}}_{#2}}

   \usepackage[english]{babel}
    \usepackage{csquotes}
    \usepackage[style=ieee, citestyle=numeric-comp,backend=biber,isbn=false,url=false,doi=false,eprint=false,maxbibnames=15]{biblatex}
\begin{document}

\title{{A QoS-Aware Uplink Spectrum and Power Allocation with Link Adaptation for Vehicular Communications in 5G networks}
}

\author{Krishna Pal Thakur, Basabdatta Palit, \textit{Member, IEEE}
\thanks{K.Pal Thakur is with the Department
of Electronics and Telecommunication Engineering, IIEST Shibpur, Howrah - 71103, West Bengal, India.}
\thanks{B.Palit is with the Department
of Electronics and Communication Engineering, NIT Rourkela, Rourkela - 769008, Odisha, India.}}
\markboth{Journal of \LaTeX\ Class Files,~Vol.~14, No.~8, August~2015}%
{Shell \MakeLowercase{\textit{et al.}}: Bare Demo of IEEEtran.cls for IEEE Journals}

\maketitle

\begin{abstract}
In this work, we have proposed link adaptation-based joint spectrum and power allocation algorithms for the uplink communication in 5G Cellular Vehicle-to-Everything (C-V2X) systems. In C-V2X, vehicle-to-vehicle (V2V) users share radio resources with vehicle-to-infrastructure (V2I) users. Existing works primarily focus on the optimal pairing of V2V and V2I users, assuming that each V2I user needs a single resource block (RB) while minimizing interference through power allocation. In contrast, in this work, we have considered that the number of RBs needed by the users is a function of their channel condition and Quality of Service (QoS) - a method called link adaptation. It effectively compensates for the frequent channel quality fluctuations at the high frequencies of 5G communication. 5G uses a multi-numerology frame structure to support diverse QoS requirements, which has also been considered in this work.

The first algorithm proposed in this article greedily allocates RBs to V2I users using link adaptation. It then uses the Hungarian algorithm to pair V2V with V2I users while minimizing interference through power allocation. The second proposed method groups RBs into resource chunks (RCs) and uses the Hungarian algorithm twice - first to allocate RCs to V2I users and then to pair V2I users with V2V users. Extensive simulations reveal that link adaptation increases the number of satisfied V2I users and their sum rate while also improving the QoS of V2I and V2V users, making it indispensable for 5G C-V2X systems.
\end{abstract}

\acresetall
 \section{Introduction}
 In recent times, \ac{C-V2X} communication has emerged as a key enabler for intelligent, automated, safe, and green vehicular communications. \ac{C-V2X} services consist of - 1) communication between vehicles and the mobile network (\ac{V2I}), and 2) direct communication between vehicles (\ac{V2V}). \ac{V2I} communication includes the exchange of road traffic and safety messages as well as other infotainment services such as voice and video calling, multimedia streaming, etc. \ac{V2V} communication, on the other hand, is mostly targeted towards improving road safety. Therefore, the \ac{QoS} requirements, such as latency and reliability, of \ac{V2V}  messages are stricter than their \ac{V2I} counterparts. For example, the \ac{BLER} requirement of \ac{V2I} users is 0.1, while that of \ac{V2V} users is 0.001~\cite{Korrai2020}. Although 5G \ac{NR} supports \ac{C-V2X} communication~\cite{Karthikeyan2020,Garcia2021}, such diverse \ac{QoS} provisioning over 5G mmWave vehicular networks is challenging due to the high frequency of operation. One possible solution is designing and implementing efficient radio resource allocation schemes for 5G \ac{C-V2X}  (NR-V2X) communication, which is addressed in this work. 
 
5G uses \ac{OFDMA}, in which the spectrum is partitioned into orthogonal time-frequency \acp{RB}. Conventionally, in 4G, all \ac{OFDMA} \acp{RB} have a uniform bandwidth and time duration. However, as services are expected to diversify further in 5G, such a uniform numerology may not suffice~\cite{Zhang2018}. Hence, to support the different \ac{QoS} requirements in 5G, 3GPP prescribes the use of multiple numerologies~\cite{TS38211,Guan2017,Son2023,Shen2022}, where each numerology is characterized by a different bandwidth and time duration of the \acp{RB}. A popular method of using these multiple numerologies is to slice the system bandwidth into orthogonal parts and assign a different numerology to each part. Users are assigned to these \acp{BWP} based on which numerology caters to their \ac{QoS} requirements. Hence, while designing resource allocation methods for 5G \ac{C-V2X} systems, it becomes imperative to account for multiple numerologies.

Another important aspect in the resource allocation in \ac{C-V2X} systems is how \ac{V2I} users share the \acp{RB} with \ac{V2V} users.  \ac{C-V2X} extensively leverages the existing  cellular network infrastructure for \ac{V2I} communications. The \ac{V2V} services, on the other hand, are provided either in  1) \textit{the underlay \ac{D2D} communication} mode~\cite{Ren2015} in which \ac{V2V} users coexist in the \ac{RB} of a \ac{V2I} user, or 2)  \textit{the dedicated mode} where \ac{V2V} users are assigned exclusive resources~\cite{Li2018, Li2019}. The former offers an efficient use of radio resources but suffers from interference between the \ac{V2I} and \ac{V2V} users. To minimize this interference, efficient power control techniques are used in the shared frequency resource, which helps to achieve a higher sum rate and satisfy the required \ac{QoS}.

 An inherent assumption in existing works on resource allocation in \ac{C-V2X} networks~\cite{Ren2015,Li2018,Li2019,Ziyan2020,He2019,Chen2019,Hou2021,Liu2023,Le2017RA,Guo2018, Guo2019,GeLi2019,Sun2016,Gyawali2019,Wu2021,Le2017,Rajan2020,Ron2022,Chai2022,Chai2023,Aslani2020,Liu2021} is that each \ac{V2I} user needs one \ac{RB}. This ignores the basic premise of radio resource allocation, which is compensating for the underlying channel impairments~\cite{Korrai2020}. Allocating a single \ac{RB} implies that all the users can use the same \ac{MCS}, and their channel condition is reasonably good. Let us consider an example. A \ac{V2I} user needs to transmit a 50-byte packet in one time slot such that the \ac{BLER} is 0.1. This packet can be sent using a single \ac{RB} of bandwidth 1440 HZ only when the modulation scheme is 16 QAM~\cite[Table 3]{Korrai2020}. Such a high modulation scheme can be achieved when the \ac{SINR} is at least as high as 11.4 dB~\cite{Korrai2020},  which may not always be achievable. For a user experiencing a lower \ac{SINR}, a lower \ac{MCS} can be supported, and, hence, multiple \ac{RB} would be needed. The resource allocation algorithm can wait for the link condition to improve until only a single \ac{RB} is required, affecting the timeliness of message delivery. On the other hand, Adaptive \ac{MCS} selection can compensate for the channel impairments - a method called Link Adaptation~\cite{Vega2021}. In this case, the choice of the \ac{MCS}, and consequently the number of \acp{RB} required, would depend on the user's underlying link condition, the amount of data to transmit at a time, and the requested \ac{QoS}. To the best of the authors'  knowledge, this aspect is amiss in existing works on multi-numerology-based resource allocation in 5G vehicular communications. So, in this work, we argue that a combination of link-adapted resource allocation, power allocation, and resource sharing in a multi-numerology setup is imperative to provide improved sum rates and QoS to different types of users in a 5G \ac{C-V2X} communication system.
 
 \indent The objective in this work is to use a  link adaptation based spectrum allocation and power allocation in the uplink to maximize the number of \ac{V2I} users and their sum rates while conforming to the \ac{QoS} guarantees of the \ac{V2I} and \ac{V2V} users. The \ac{QoS} guarantees are specified by outage probability constraints, average packet delays, and \ac{BLER} requirements. Due to link adaptation, the number of \acp{RB} allocated to each user becomes a variable quantity, thereby making the resource sharing between \ac{V2I} and \ac{V2V} users more challenging. In addition, we have also considered the presence of multiple best-effort users along with \ac{V2I} and \ac{V2V} users. To support such diverse \ac{QoS} requirements, we have used multiple numerologies~\cite{TS38211}. We have, thus, reformulated the problem of spectrum and power allocation in \ac{C-V2X} communication for 5G networks, wherein we have not only considered the pairing between \ac{V2I} users and underlaid \ac{V2V} users but also the link adaptation based resource allocation in a multi-numerology setup. The proposed method offers a more comprehensive and practical approach towards resource allocation in 5G \ac{C-V2X}. The major contributions are:
 \begin{enumerate}
     \item We have proposed two resource allocation algorithms for \ac{V2I} and \ac{V2V} users in the shared mode.
     \begin{enumerate}
         \item {The first algorithm, \textit{Greedy Resource Allocation with Hungarian Sharing (GRAHS)} operates as follows. - (i) First, it greedily allocates \acp{RB} to \ac{V2I} users through an adaptive selection of \acp{MCS}. (ii) It then allocates power to all the \ac{V2I} and \ac{V2V} user pairs, in the \acp{RB} that are assigned to the respective \ac{V2I} user in step (i), such that the \ac{V2I} datarate is maximized. (iii) Using the new power allocation from step (ii), GRAHS pairs the \ac{V2I} and \ac{V2V} users using the Hungarian algorithm such that the respective QoS requirements and outage probability conditions are satisfied~\cite{Le2017RA,Sun2016,West2000}.
         }
         \item As GRAHS allocates non-contiguous resources, hence we have also proposed a second algorithm, \textit{called Hungarian Resource Allocation with Hungarian sharing (HRAHS)},  to maximize the \ac{V2I} datarate. This algorithm groups a set of  \acp{RB} as a \ac{RC} \cite{Calabrese2008,Mukho2015} and assigns these \acp{RC} to the \ac{V2I} users using the Hungarian algorithm. This is followed by another Hungarian algorithm, which pairs the scheduled \ac{V2I} and the \ac{V2V} users.
    \end{enumerate}
     \item We have also explored the dedicated or overlay mode ~\cite{Li2018, Li2019} in which \ac{V2I} and the \ac{V2V} users are assigned orthogonal resources. We have compared the three algorithms - GRAHS, HRAHS, dedicated mode of resource allocation - in terms of their real-time traffic capacity and sum rate. We have defined the real-time traffic capacity as the number of \ac{V2I} users for which at least 95\% users have -1) a packet loss rate less than 2\%, and 2) average packet delay less than the delay bound~\cite{Palit2015}.
 \end{enumerate}
 Extensive simulations and comparison with an existing work~\cite{Le2017} show that link adaptation increases the number of supported \ac{V2I} users by nearly five times when it is used in dedicated mode and seven times when link adaptation is used in the shared mode.\\
 \indent The rest of the article is organized as follows. Section \ref{sec:RelWork} discusses the related work. Section \ref{sec:Sys_Model} explains the system model and problem formulation. Section \ref{sec:Methodology} describes the proposed algorithms. Section \ref{sec:Results} discusses the results obtained. Section \ref{sec:Conclusion} concludes the article.
\section{Related Work}\label{sec:RelWork}
\small
\begin{table*}[t]
\centering
\caption{Summary of Recent Works on Spectrum Allocation in \ac{C-V2X} Networks and Comparison with the Current Work
}\label{tab:summary}
\begin{tabular}{|p{1.15cm}|p{12cm}|l|l|l|l|l|l|}
\hline
\textbf{References} &  \textbf{Summary} & \begin{sideways} \textbf{Power Allocation}\end{sideways} & \begin{sideways} \textbf{Resource Allocation} \end{sideways} &  \begin{sideways} \textbf{Resource Sharing}\end{sideways}&\begin{sideways} \textbf{QoS Awareness}\end{sideways}&\begin{sideways} \textbf{Imperfect CSI}\end{sideways}&\begin{sideways} \textbf{Link Adaptation}\end{sideways}\\
\hline
~\cite{Ren2015}           & Proposes a location information-aided power control mechanism to reduce interference in the reused channel for V2V communication over underlaid D2D links.  & \Checkmark & $\times$    & \Checkmark  &  $\times$ & $\times$ & $\times$\\
  \hline
~\cite{Li2018}           & Studies three resource allocation modes for VUEs - eNB assisted dedicated and shared modes, and distributed or autonomous mode; proposes a joint resource mode selection and resource allocation scheme. & $\times$ & \Checkmark  &  \Checkmark  &  $\times$ &  $\times$ & $\times$\\
  \hline  
  ~\cite{Li2019}           & Extends~\cite{Li2018} to include power allocation to reduce interference between V2V users and PUEs.  & \Checkmark & \Checkmark   & \Checkmark  &  $\times$ &  $\times$ & $\times$\\
  \hline 
    ~\cite{Ziyan2020}           & Proposes a resource allocation algorithm that uses a reinforcement learning-based GNN framework for resource sharing between \ac{V2V} and \ac{V2I} users to maximize the sum-capacity. & $\times$ & \Checkmark   & \Checkmark  &  $\times$ & $\times$ &  $\times$\\
      \hline 
    ~\cite{He2019}           & Proposes a resource allocation scheme in which one \ac{V2I} link is shared by multiple \acp{V2V} users. \ac{V2V} pairs are clustered using coalition games, and power allocation is used to minimize interference.  & \Checkmark & $\times$  & \Checkmark  &  $\times$ &  $\times$ & $\times$\\
       \hline 
           ~\cite{Chen2019}           &
    Proposes a power allocation and a belief propagation-based \ac{RB} allocation in which multiple \ac{V2V} links share \acp{RB}; \acp{RB} are allocated to maximize sum capacity.  & \Checkmark & \Checkmark  & \Checkmark  &  $\times$ &  $\times$ & $\times$\\
 \hline
    ~\cite{Hou2021}           & 
    Proposes a deep learning-based spectrum reuse and power allocation scheme. & \Checkmark & $\times$  & \Checkmark  &  $\times$ & $\times$ & $\times$\\
    \hline 
       ~\cite{Le2017RA}           & 
Proposes a power allocation and spectrum sharing method for coexisting \ac{V2I} and \ac{V2V} users to maximize the sum-ergodic capacity and the minimum ergodic capacity of the \ac{V2I} users while satisfying the minimum link guarantees for \ac{V2V} users; uses the Hungarian algorithm to pair \ac{V2I} and \ac{V2V} users. & \Checkmark & $\times$ & \Checkmark  &  \Checkmark &  $\times$ & $\times$\\
    \hline 
      ~\cite{Guo2018}           & Proposes a power allocation and spectrum reuse policy that taps large-scale fading information to maximize sum ergodic capacity of \ac{V2V} and \ac{V2I} users; analyses packet sojourn time.  & \Checkmark & $\times$  & \Checkmark  &  \Checkmark & $\times$ & $\times$\\
    \hline 
      ~\cite{Guo2019}           & 
Uses the theory of effective capacity on the large scale fading information to propose a joint power allocation and spectrum reuse method which maximizes the sum ergodic capacity within the constraints of the delay violation probability for \ac{V2I} and \ac{V2V} users. & \Checkmark & $\times$  &   \Checkmark & \Checkmark & $\times$ & $\times$\\
    \hline         
       ~\cite{GeLi2019}           & 
Extends the work in~\cite{Guo2018} to include packet retransmission so that the reliability of the transmission can be increased; analyses the packet delay and packet sojourn time using queueing theory.& \Checkmark & $\times$  & \Checkmark  &  \Checkmark &  $\times$ & $\times$\\
    \hline     
      ~\cite{Sun2016} & Proposes a heuristic method for non-orthogonal resource sharing and power allocation between \ac{V2I} and \acp{V2V} users, considering the availability of slowly varying CSI while guaranteeing latency and reliability.  & \Checkmark & $\times$  & \Checkmark  &  \Checkmark &  $\times$ & $\times$\\
    \hline 
     ~\cite{Gyawali2019}           & Considers a limited availability of accurate CSI; Uses centralized Hungarian algorithm to assign resources to \ac{V2I} users and distributed DRL for power control and resource sharing between \ac{V2I} and \ac{V2V} users.  & \Checkmark & \Checkmark  & \Checkmark  &  $\times$ &  \Checkmark & $\times$\\
    \hline 
         ~\cite{Wu2021}           & Uses probabilistic information on CSI for resource allocation of \ac{V2I} users, power control and subsequent resource sharing between \ac{V2I} and \ac{V2V} users.  & \Checkmark & $\times$  & \Checkmark  &  \Checkmark & \Checkmark & $\times$\\
    \hline 
      ~\cite{Le2017}           & 
Extends the work in~\cite{Le2017RA} to include high mobility conditions where imperfect CSI feedback is available; maximizes the sum capacity while maintaining the link guarantees for V2V users.  & \Checkmark & $\times$  & \Checkmark  &  \Checkmark &  \Checkmark & $\times$\\
\hline
~\cite{Rajan2020}        & 
Proposes to maximize the sum ergodic capacity of all \ac{V2V} pairs subject to a minimum \ac{SINR} requirement of \ac{V2I} links using a low complexity power allocation approach and the Gale Shapley method to pair multiple \ac{V2V} links with one \ac{V2I} links.  & \Checkmark & $\times$  & \Checkmark  &  \Checkmark &  \Checkmark & $\times$\\
\hline
~\cite{Aslani2020}        & 
Uses large-scale fading information to propose a power control and subcarrier assignment algorithm for \ac{V2V} and \ac{V2I} users such that the number of \ac{V2V} links is maximized subject to the diverse \ac{QoS} requirements of \ac{V2I} and \ac{V2V} users.  & \Checkmark & \Checkmark  & \Checkmark  &  \Checkmark &  $\times$ & $\times$\\
\hline
 This work           & 
Proposes link adapted resource allocation for \ac{V2I} users and a joint power allocation and resource sharing for \ac{V2I} and \ac{V2V} users to increase the number of supported \ac{V2I} users and their sum capacity while conforming to the QoS guarantees of \ac{V2I} and \ac{V2V} users.  & \Checkmark & \Checkmark  & \Checkmark  &  \Checkmark &  \Checkmark & \Checkmark\\
    \hline
\end{tabular}
\end{table*}
\begin{table*}[t]
\centering
\caption{Symbol Definitions}\label{tab:symbdef}
\begin{tabular}{|p{1cm}|p{6cm}|p{1cm}|p{6cm}| } 
 \hline
 \textbf{Symbol} & \textbf{Definition} & \textbf{Symbol} & \textbf{Definition} \\ 
 \hline 
 $c$ & index of Cellular User-Equipments (CUEs) & $\gamma^n_i$ & SINR of UE $i$ in the RB $n$, $i\in\{c,v\}$ \\
 \hline
 $v$ & index of Vehicular User-Equipments (VUEs)& $R_c$ & Rate of CUE $c$ \\
 \hline
 $m$ & index of Best-effort User-Equipments (BUEs)& $\rho^n_c$ & Resource Block allocation variable\\ \hline
  $\delta_i$& Time-to-live of packets of user type $i$, $i\in\{c,v\}$ &  $x_{c,v}$ & Resource Sharing indicator variable\\ 
 \hline 
  $\beta_i$ & Packet Size of user type $i\in\{c,v\}$ & $\eta^j_c$ & Resource Chunk Allocation variable for HRAHS  \\ \hline
  $\mathcal{B}_i$  & Time-Domain buffer  of user type $i\in\{c,v\}$  &  $D_i$ & Average packet delay of user type $i\in\{c,v\}$\\
 \hline

 $\epsilon$&Channel Correlation Coefficient & $r_0$ & Minimum rate requirement of CUEs\\
 \hline
 \
 $P^\mathcal{C}_i$ & Transmit power of user type $i\in\{c,v\}$ & $p_0$ & Maximum Outage probability of VUEs \\
 \hline
 $P^\mathcal{C}_{max}$ & Maximum transmit power of CUEs & $\gamma_0$ & Minimum SINR Threshold for  VUEs\\ \hline
 $P^\mathcal{V}_{max}$ & Maximum transmit power of {VUEs} &  $C_t$ & maximum number of users  scheduled  per TTI\\ \hline
  $N_v$ & Number of RBs required by a VUE $v$  & $N_c$ & Number of RBs required by CUE $c$\\  \hline
 
\end{tabular}
\end{table*}
\normalsize
 \indent This section discusses the existing works on resource and power allocation in \ac{C-V2X} systems. These works primarily consider three aspects: (i) spectrum allocation, (ii) spectrum sharing, and (iii) power allocation for interference minimization in shared resources.\\ 
\indent The work in~\cite{Ren2015} proposes a location-information-aided power control technique for  D2D-based vehicular communication networks where multiple V2V users share the link of a \ac{V2I} user in the uplink. The algorithm in~\cite{He2019} allocates power using the difference of two convex functions approach and uses a game theory for V2V clustering. Authors in~\cite{Liu2023} propose a \ac{DRL} based control aware power and radio resource allocation algorithm for vehicle platoons. The work in~\cite{Li2018} considers a mixed mode of resource allocation, in which \ac{V2V} users either - (i) get dedicated \acp{RB}, (ii) or share \acp{RB} with the \ac{PUE} which directly communicate with the \ac{BS}, (iii) or select the \acp{RB} themselves without any involvement of the \ac{eNB}. Authors in~\cite{Li2019} extend the work in~\cite{Li2018} to include a power allocation for interference control and a low-complexity user-pairing method. The resource allocation algorithm in~\cite{Ziyan2020} uses reinforcement learning in graph neural networks to allocate resources to \ac{V2V} and \ac{V2I} users while maximizing the sum capacity. The deep learning-based algorithm in~\cite{Chen2019} takes decisions on resource sharing between \ac{V2I} and \acp{V2V} users. The work in~\cite{Hou2021} explores belief propagation and a factor graph model to investigate if multiple \ac{V2V} links can share the \ac{RB} assigned to a \ac{V2I} user. However,~\cite{Ren2015,Ziyan2020,He2019,Chen2019,Hou2021,Li2019,Liu2023} need full and perfect \ac{CSI} for power and resource allocation.  However, the high mobility-induced fast channel strength variations cause Doppler Spread, delaying instantaneous \ac{CSI} feedback.\\
\indent The need for complete knowledge of the instantaneous \ac{CSI} of the \ac{V2V} users is bypassed in \cite{Le2017RA,Guo2018, Guo2019,GeLi2019,Ji2023} by using the large-scale fading information. The algorithm in \cite{Le2017RA} proposes to allocate power and share spectrum based on the large-scale fading to maximize the sum capacity of \ac{V2I} links while ensuring a minimum coverage probability of \ac{V2V} users. In~\cite{Guo2019}, the effective capacity theory is applied to large-scale fading to devise a power allocation and spectrum reuse scheme. The work in ~\cite{Ji2023} proposes a power allocation and spectrum sharing algorithm, which applies multi-agent reinforcement learning on large-scale fading information to maximize sum-capacity of \ac{V2I} users and provide \ac{QoS} guarantees and secrecy outage guarantees to \ac{V2V} users. Large-scale fading depends only on user location; hence, the changes in channel gain vary slowly with time. The work in \cite{Sun2016} uses a slowly varying \ac{CSI}, instead of large-scale fading, for non-orthogonal resource sharing and power allocation between \ac{V2I} and \ac{V2V} users. In ~\cite{Guo2018}, the algorithm maximizes the sum rate of \ac{V2I} users while satisfying the delay requirement of \ac{V2V} users, assuming that the \ac{BS} knows only the distribution of the small-scale channel fading. The work in~\cite{Guo2018} is extended in \cite{GeLi2019}  to include packet retransmission. On the other hand, ~\cite{Wu2021} uses the probabilistic information instead of the accurate \ac{CSI} to allocate radio resources. \\
\indent Unlike the aforementioned works, imperfect CSI is considered in the works~\cite{Gyawali2019, Le2017,Rajan2020,Ron2022,Chai2022,Chai2023,Aslani2020,Liu2021}. The work in~\cite{Le2017} extends the work in~\cite{Le2017RA} wherein the joint spectrum and power allocation algorithm considers delayed \ac{CSI} feedback.The work in~\cite{Gyawali2019} proposes a combination of centralized and distributed resource allocation based on graphs and reinforcement learning. The proposed algorithm uses limited \ac{CSI} to allocate spectrum to the \ac{V2I} users, and the \ac{V2V} users exploit the realistic \ac{CSI} for power control. Authors in~\cite{Rajan2020} propose a one-to-many matching resource sharing problem where one \ac{V2V} pair can share the \acp{RB} assigned to multiples \ac{V2I} users. In ~\cite{Ron2022} is proposed an unsupervised learning-based power control technique that maximizes the sum capacity of \ac{V2I} and \ac{V2V} users. Data-driven methods are used in~\cite{Chai2022, Chai2023} for resource allocation and \ac{V2V} user grouping with partial-\ac{CSI} in \ac{C-V2X} networks. The work in~\cite{Aslani2020} addresses the problem of heterogeneous \ac{QoS} provisioning through joint power and spectrum allocation while trying to maximize the number of \ac{V2V} pairs. It assumes that the global \ac{CSI} information is unavailable.\\
 \indent In contrast to existing works, in this article, we have not only considered resource sharing and power allocation but have also incorporated the link-adapted resource allocation of \ac{V2I} users to maximize the \ac{V2I} user capacity as well as their sum capacity, while supporting the stringent delay and \ac{QoS} requirements of both \ac{V2I} and \ac{V2V} users.

\section{System Model}\label{sec:Sys_Model}
\begin{figure}[t]
\centering
\includegraphics[trim = {0cm 0.5cm 0cm 0.5cm}, width = 0.49\textwidth]{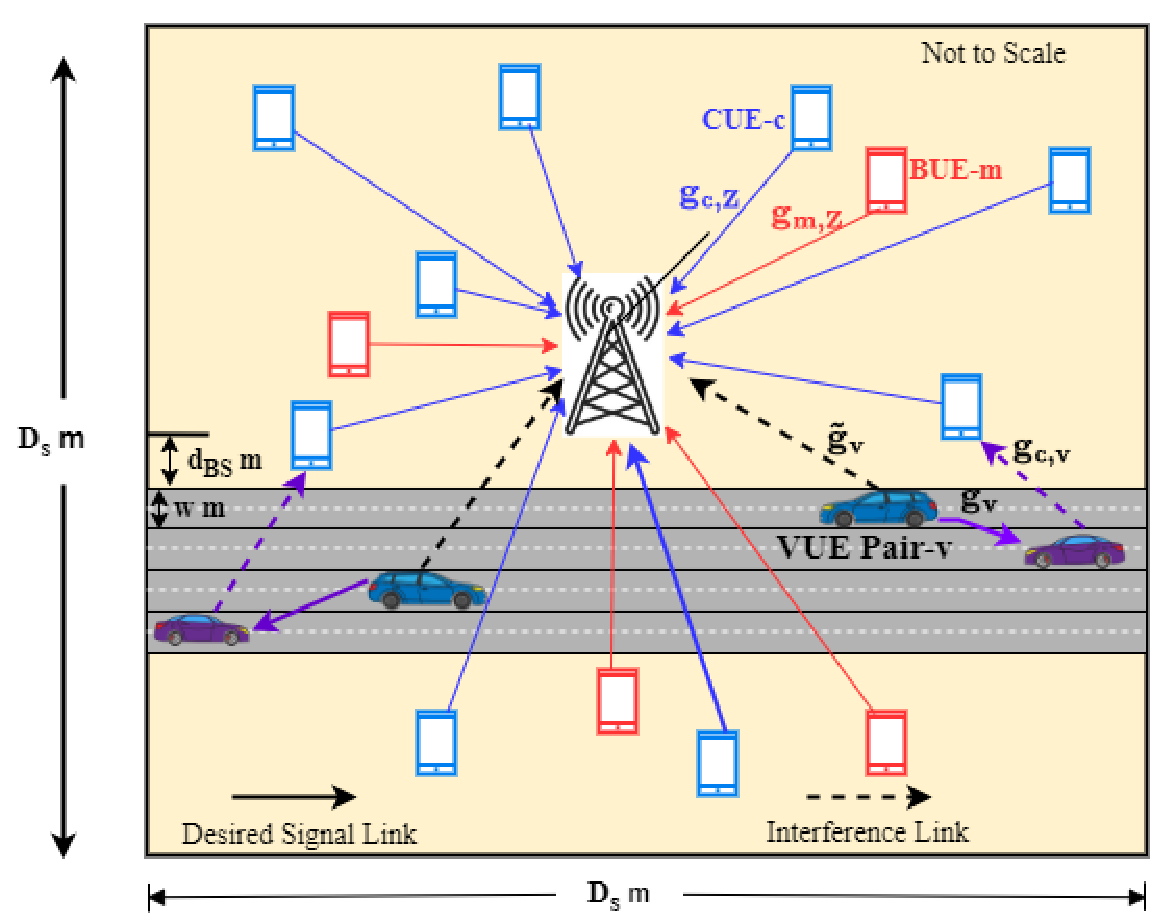}
\caption{A simplified version of the System Model. Only a few CUEs, BUEs, VUE pairs, and four lanes have been shown in the \ac{gNB} service area.}
\label{fig:SM}
\end{figure}

Fig. \ref{fig:SM} shows an overview of the system model, developed according to \cite{TS38211}, and is explained in this section along with the problem formulation. Table \ref{tab:symbdef} outlines the symbol definitions.
\subsection{System Model}\label{sec:sys_model}
  \subsubsection{\textbf{Scenario and User Distribution}}\label{sec:scenario}
  {In this work, we have considered the service area of a single \ac{gNB}, in which the \ac{gNB} is located at the center as shown in Fig.~\ref{fig:SM}.
  We have assumed that there are $N_L$ lanes inside this area, each of width $w$ metres, located to the south of the \ac{gNB}.}
   There are $C$ stationary \ac{V2I} users or \acp{CUE}, $V$ \ac{V2V} pairs or  \acp{VUE}, and $M$ stationary \acp{BUE}. The \acp{CUE} and \acp{BUE} are uniformly distributed in the service area, excluding the lanes. All the \acp{VUE} have  the same velocity $\nu$ and have been dropped uniformly in the lanes~\cite{TR36885}.
 We have shown in Fig.\ref{fig:SM} only a few \ac{CUE}, \acp{BUE}, and \ac{VUE} pairs and four lanes to maintain clarity.
\subsubsection{\textbf{User Traffic and Packet Generation}}\label{sec:UTandPG}
 The \acp{CUE} and \acp{BUE} communicate directly with the \ac{gNB}. We have assumed that the network traffic generated by the \acp{CUE} is mostly of infotainment nature, such as voice or video calling, multimedia streaming, etc. These users, thus, have a rate constraint and 
a delay constraint. The transmitter and the receiver of the \ac{VUE} pairs communicate only amongst themselves and primarily exchange safety enhancement messages. So, they have a coverage probability constraint, and their delay constraint is more stringent than the \acp{CUE}. Thus, the \acp{CUE} and \acp{VUE} generate delay-sensitive real-time traffic. While a \ac{CUE} generates a packet of $\beta_c$ bytes every $\tau_c$ ms, a \ac{VUE} generates a packet of $\beta_v$ bytes  every $\tau_v$ ms. The number of \acp{CUE} generating a packet in a time slot follows a Poisson distribution with a mean $\lambda_c = \frac{\text{Total no. of \acp{CUE}}}{20\text{ms}   }$.    The delay constraint or time-to-live for the \ac{CUE} and \ac{VUE} packets is $\delta_c$ ms and $\delta_v$ ms, respectively. The \acp{BUE} essentially communicate non-real-time traffic with no \ac{QoS} requirement. Their traffic is of full buffer type, \textit{i.e.} they always have some data to send.  
\subsubsection{\textbf{Multiple Numerologies and OFDMA}}\label{sec:multinumerology}
\begin{table}[]
\centering
\caption{Different Numerologies in 5G NR}\label{tab:numerology}
\begin{tabular}{|c|c|c|c|}
\hline
\begin{tabular}[c]{@{}c@{}}\textbf{Numerology} \\\textbf{ Index} \\ $\mu$\end{tabular} & \begin{tabular}[c]{@{}c@{}}\textbf{Sub-Carrier} \\ \textbf{Spacing}\\ $\Delta f = 2^\mu\times\mathrm{15 KHz} $\end{tabular} & \begin{tabular}[c]{@{}c@{}}\textbf{Bandwidth} \\ \textbf{of one RB} \\ (\textbf{KHz})\end{tabular} & \begin{tabular}[c]{@{}c@{}}\textbf{TTI}  \\ \textbf{duration} \\ \textbf{(ms)}\end{tabular} \\ \hline
0 & 15 & 180 & 1 \\ \hline
1 & 30 & 360 & 0.5 \\ \hline
2 & 60 & 720 & 0.25 \\ \hline
3 & 120 & 1440 & 0.125 \\ \hline
4 & 240 & 2880 & 0.0625 \\ \hline
\end{tabular}
\end{table}
\indent To support the diverse \ac{QoS} requirements as mentioned in Section \ref{sec:UTandPG}, we have used the multiple numerology frame structure of 5G NR~\cite{TS38211}. We have divided the system bandwidth into two bandwidth parts (\acp{BWP}), each supporting different numerologies. As 5G uses \ac{OFDMA}, each \ac{BWP} is divided into orthogonal time-frequency radio \acp{RB} spanning 12 \ac{OFDM} sub-carriers in the frequency domain and 14 OFDM symbols in the time domain. So, the bandwidth of a \ac{RB} is $12\times\mathrm{\ac{SCS}}$, and the duration of 14 symbols constitutes one \ac{TTI}.  The different numerologies are characterized by  different \acp{SCS} and different  \ac{TTI} durations, as specified in Table \ref{tab:numerology}. Users of different applications are assigned different numerologies, \textit{i.e.}, to different BWPs based on their \ac{QoS} requirements. For example, the numerology in \ac{BWP}-1 is $\mu=3$ with a  \ac{SCS} of $120$ KHz and a  TTI duration of $0.125$ milliseconds (ms). It caters to time-critical applications, such as the \acp{CUE} and \acp{VUE}. On the other hand, \ac{BWP}-2 of bandwidth $B_2$ has a numerology $\mu = 0$ with a \ac{SCS} of $15$ KHz and a slot duration of $1$ ms, and it serves the \acp{BUE}. \\ 
 \indent To minimize the interference among the neighbouring \acp{CUE}, the \ac{gNB} allocates orthogonal uplink \acp{RB} from \ac{BWP}-1 to the \acp{CUE}. $\rho^n_c$ represents the allocation variable, such that $\rho^n_c =  1$, if \ac{RB} $n$ is allocated to \ac{CUE} $c$, and $\rho^n_c =  0$, otherwise.
 \subsubsection{\textbf{Modes of RB Allocation}}\label{sec:modeRB}
 \indent We explore two modes of \ac{RB} allocation between the \acp{CUE} and \acp{VUE}~\cite{Li2019} -- 
\begin{enumerate}[(i)]
    \item \textbf{Shared or Underlay mode} - in which the \acp{VUE} operate in the \ac{D2D} side link, \textit{i.e.}, the uplink resources of  CUEs are reused by the VUEs. This is made possible because of the small amount of interference that \acp{VUE} generates to the uplink communication between the \acp{CUE} and the \ac{gNB}. Power control can minimize this interference, as explained in Section \ref{sec:power_alloc}. We set the indicator variable $x_{c,v}=1$ if the \acp{RB} of \ac{CUE} $c$ are shared with  \ac{VUE} $v$, otherwise we set, $x_{c,v}=0$, $\forall c \in \mathcal{C} = \left\{1, 2, \cdots, C\right\}$, and $\forall v \in \mathcal{V} = \left\{1, 2, \cdots, V\right\}$.
    \item \textbf{Dedicated or Overlay Mode} - in which the \acp{CUE} and the \acp{VUE} are assigned orthogonal \acp{RB}.
\end{enumerate} 
 

\subsubsection{\textbf{Channel Modeling}}
\indent The next important step is channel modeling. We assume that the small-scale fast-fading gain $h^n_v$ of the \ac{VUE} $v$ in RB-$n$, is independent and identically distributed (i.i.d) as $\mathcal{CN}(0,1)$ for all \acp{VUE}. We further assume a block fading channel model in which $h^n_v$ for \ac{VUE} $v$ is different in different \acp{RB} in the same \ac{TTI}. The corresponding channel gain $g^n_v$ of the \ac{VUE} pair $v$ in \ac{RB}-$n$ is given by, \begin{equation}
  g^n_v = |h^n_{v}|^2\alpha_{v}, \label{eq:CSI}
\end{equation}
where $\alpha_{v}$ is the large-scale fading gain. Applying the assumptions mentioned above to the different communication links in Fig. \ref{fig:SM}, the respective channel gains can be defined as: 
\begin{enumerate}
    \item $g^n_{c,Z}$ - Uplink channel gain in \ac{RB}-$n$ between a \ac{CUE} $c$ and the \ac{gNB} $Z$,
    \item $g^n_{m,Z}$ - Uplink channel gain in \ac{RB}-$n$ between a \ac{BUE} $m$ and the \ac{gNB} $Z$,
    \item $g^n_v$ - channel gain of a \ac{VUE} pair over \ac{RB}-$n$,
    \item $\Tilde{g^n_v}$ - Interference from \ac{VUE} pair $v$ to the \ac{CUE}-\ac{gNB} communication over \ac{RB}-$n$,
    \item $\Tilde{g}^n_{c,v}$ - Interference from the   \ac{CUE} $c$ to the  \ac{VUE} $v$ in \ac{RB}-$n$.
\end{enumerate}

 Large-scale fading, composed of path loss and shadowing, depends only on the user positions and varies slowly. Hence, they are perfectly determined at the \ac{gNB} for all links. It is also assumed that the channel gains, $g^n_{c,Z}$, $g^n_{m,Z}$ and $\Tilde{g^n_v}$, of links directly connected to the \ac{gNB}  can be  
accurately obtained at the \ac{gNB}. However,  the channel gains, ${g^n_v}$ and $g^n_{c,v}$, of links that are not directly connected to the \ac{gNB} are reported with a feedback period of one TTI, of duration $T$. The high mobility of vehicles also introduces the Doppler shift in the small-scale fading. We, thus, assume that the gNB
can only obtain the estimated channel fading $h$ with an error $e$. The gain $h$ follows a first-order Gauss-Markov process over a period $T$, \textit{i.e.},
\begin{align}
h=\epsilon \hat{h} + e,
\end{align}
where $h$ and $\hat{h}$ represent the small-scale fading gain in the current and previous \acp{TTI}, respectively. The channel difference, $e$, between two consecutive \acp{TTI}, follows an i.i.d $\mathcal{CN}(0, 1-\epsilon^2)$  distribution. It is independent of $\hat{h}$. The channel correlation coefficient $\epsilon$ between  two consecutive \acp{TTI} follows Jake's fading model~\cite{Stuber1996}, such that $\epsilon = J_0(2\pi f_d T)$. Here $J_0(.)$ is the zeroth-order Bessel function of the first kind, $f_d = \nu f_c/c$ is the maximum Doppler shift in frequency with $c = 3 \times 10^8$ m/s, $\nu$ is the vehicular speed, and $f_c$ is the carrier frequency.\\
\begin{table*}[t]
\centering
\caption{Modulation and Coding Schemes, Spectral Efficiency, and Signal-to-Noise Ratio ranges to support different \acp{BLER}}
\label{tab:MCS}
\begin{tblr}{
  cells = {c},
  hlines = {0.05em},
  vline{-} = {1-7}{},
  hline{1,7} = {-}{0.08em},
}
\textbf{MCS}                                     & 1    & 2    & 3    & 4    & 5    & 6    & 7          & 8          & 9          & 10         & 11         & 12         & 13         & 14         & 15         \\
\textbf{Modulation}                              & QPSK & QPSK & QPSK & QPSK & QPSK & QPSK & {16\\ QAM} & {16\\ QAM} & {16\\ QAM} & {64\\ QAM} & {64\\ QAM} & {64\\ QAM} & {64\\ QAM} & {64\\ QAM} & {64\\ QAM} \\
{\textbf{Spectral Efficiency} \\ \textbf{(bits/symbol)}}                   & 0.15 & 0.23 & 0.38 & 0.6  & 0.88 & 1.18 & 1.48       & 1.91       & 2.41       & 2.73       & 3.32       & 3.90       & 4.52       & 5.12       & 5.55       \\
{\textbf{SNR Th (dB)}\\$\mathrm{BLER_c} = 0.1$}  & -6.5 & -4.0 & -2.6 & -1   & 1    & 3    & 6.6        & 10.0       & 11.4       & 11.8       & 13.0       & 13.8       & 15.6       & 16.8       & 17.6       \\
{\textbf{SNR Th (dB)}\\$\mathrm{BLER_v} = 0.01$} & -2.5 & 0.0  & 1.4  & 3.0  & 5.0  & 7.0  & 10.6       & 14.0       & 15.4       & 15.8       & 17.0       & 17.8       & 19.6       & 20.8       & 21.6       \\
{\textbf{No. of RBs}\\ \textbf{required for a} \\\textbf{50 byte packet}}        & 16   & 11   & 7    & 4    & 3    & 3    & 2          & 2          & 1          & 1          & 1          & 1          & 1          & 1          & 1          \\
{\textbf{No. of RBs}\\ \textbf{required for a} \\\textbf{10 byte packet}}        &4    &3    &2     & 1    & 1    &1    &1          &1           & 1          &1           &1           &1           &1           &1           &1           
\end{tblr}
\end{table*}
\indent 
The \ac{SINR} of the  \ac{CUE} $c$ in the \ac{RB}-$n$ in shared mode is,
\begin{equation}
  \gamma^n_c =\frac{ \po{C}{c} \alpha_{c,Z}|h^n_{c,Z}|^2}  {\sigma^2+\sum_{v=1}^{V} x_{c,v}\po{V}{v} \Tilde{\alpha}_v|\Tilde{h}^n_v|^2}, 
  \label{eq:SINR_CUE}
 \end{equation}
and the \ac{SINR} of the \ac{VUE} $v$ is.
\begin{equation}
\gamma^n_v =\frac{\po{V}{v} \alpha_{v}(\epsilon_v^2 |\hat{h}^n_v|^2 + |e^n_v|^2)}{\sigma^2+\sum_{c=1}^{C} x^n_{c,v} \po{C}{c} (\epsilon_{c,v}^2 |\hat{h}^n_{c,v}|^2 + |e^n_{c,v}|^2) }, 
\label{eq:SINRVUE}
\end{equation}
where $\po{C}{c}$ and $\po{V}{v}$ are the transmit powers of \ac{CUE} $c$ and \ac{VUE} $v$, respectively. $\sigma^2$  is the noise power. We assume a packet is served within one \ac{TTI} without any retransmission. 
\subsubsection{\textbf{Link Adaptation}}\label{sec:LA}
{Finally, we discuss link adaptation, which is used in this work to tune the number of \acp{RB} required by a packet to the underlying channel conditions. This is in contrast to existing works that implement \ac{CUE} packet transmission using a single \ac{RB} only. The number of \acp{RB} required to transmit a packet in a TTI is calculated as {$\mathtt{\frac{Packet\ size\ in\ bits}{Num\ of\ bits/ RB}}$, where $\mathtt{Num\ of\ bits/ RB =  num\ of\ symbols/ RB \times num\ of\ bits/symbol}$}. The number of symbols per RB depends on the number of sub-carriers in a \ac{RB} and the number of symbols in a \ac{TTI}. So, from Section \ref{sec:multinumerology}, the
number of symbols per RB is $\mathtt{12 \times 14}$, and the number of bits per RB is $\mathtt{12\ \times 14\ \times number\ of\ bits/symbol}$. The number of bits per symbol, i.e.,  the user's  \ac{SE}, on the other hand, depends on the \ac{MCS}, which in turn depends on the user's \ac{SINR} and \ac{BLER} requirement. Table \ref{tab:MCS}~\cite[Table 3]{Korrai2020} shows the different \acp{MCS}, the corresponding \ac{SE} and \ac{SINR} ranges for different \acp{BLER}. It may be observed that for a given \ac{BLER}, a higher \ac{SINR} supports a higher \ac{MCS}, and, hence, a higher \ac{SE}. However, for the same \ac{SINR}, a lower \ac{BLER} implies a lower \ac{MCS} needs to be selected. Table ~\ref{tab:MCS} also shows the number of \acp{RB} needed to serve a \ac{CUE} or a \ac{VUE} packet. So, once the \ac{SINR} of the \ac{CUE} (or \ac{VUE}) is obtained, its MCS and, hence, its \ac{SE} for a given \ac{BLER} is obtained from Table \ref{tab:MCS}. Subsequently, the number of \acp{RB} required is also calculated according to the packet size.}

\subsection{Problem Formulation for Shared mode}
This section explains the problem formulation for the shared resource allocation mode. We aim to find a) the optimal resource allocation of \acp{CUE}, and 2) the optimal pairing between \acp{CUE}-\acp{VUE} such that the sum rate of the \acp{CUE} is maximized. At the same time, the delay and rate constraints of  \acp{CUE} and the delay and coverage probability constraints of  \acp{VUE} should be satisfied. 
We denote the transmission rate of \ac{CUE}-$c$ in \ac{RB}-$n$ is $R_c^n$, which can be obtained from Table \ref{tab:MCS}. So, the optimization problem is:
\begin{align}
    {\max_{\{\rho^n_c\},\{x_{c,v}\}, \{\po{V}{v}\},\{\po{C}{c}\}} \sum_n\sum_{c\in\mathcal{C}} R^n_c \rho^n_c }
 \label{eq:Sumrate}
\end{align}
s.t.
\begin{subequations}
\begin{align}
    &{\sum_{c\in\mathcal{C}}\rho^n_c \leq 1,\; \forall n, 
    \label{eq:RAconstraint}}\\
     &D_c \leq \delta_c, \; \forall c \,\in\, \mathcal{C};\;
    D_v \leq \delta_v, \; \forall v \,\in\, \mathcal{V} \label{eq:veh_delay_const}\\
    &\sum_n R^n_c \rho^n_c \geq r_0,\; \forall c \,\epsilon\, \mathcal{C}\label{eq:rate_const}\\ 
    &\mathtt{Pr}\{ \sum_n\gamma^n_v\rho^c_n x_{c,v} \leq \gamma_0\} \leq p_0,\; \forall v \in \mathcal{V}\label{eq:out_p_const}\\
     &0\leq \po{C}{c} \leq P^\mathcal{C}_\mathrm{max}, \; \forall c \in \mathcal{C}\label{eq:cell_power_const}\\
    &0\leq \po{V}{v} \leq P^\mathcal{V}_\mathrm{max}, \; \forall v \in \mathcal{V}\label{eq:veh_power_const}\\
    &\sum_{c\epsilon \mathcal{C}} x_{c,v} \leq 1,\; x_{c,v} \, \in \, \{0,1\},\; \forall v \in \mathcal{V}\label{eq:resource_sharing_const_1}\\
    &\sum_{v\epsilon V} x_{c,v} \leq 1,\; x_{c,v} \, \in \, \{0,1\},\; \forall c \,\in\, \mathcal{C}\label{eq:resource_sharing_const_2}.
\end{align}
\end{subequations}
{
The details of the constraints are as follows. 
\begin{itemize}
    \item  \textit{\textbf{\ac{RB} allocation constraint}} - (\ref{eq:RAconstraint}) says that (i) one \ac{RB} can be assigned to only one \ac{CUE}. The total number of \acp{RB} occupied by the \acp{CUE} should be less than or equal to $N_1$, i.e., the number of \acp{RB} in \ac{BWP}-1.
    \item \textbf{\textit{Delay constraint}} - (\ref{eq:veh_delay_const})  says that the average packet delays, $D_c$ and $D_v$ of a \ac{CUE} and \ac{VUE} should be less than their time-to-live $\delta_c$ and $\delta_v$, respectively. 
    \item \textbf{\textit{Minimum Rate constraint}} - (\ref{eq:rate_const}) says that \acp{RB} are shared between \ac{CUE} and \ac{VUE} if  the minimum required rate $r_0$ of \acp{CUE} is guaranteed.  
    \item \textbf{\textit{Coverage Probability constraint}} - (\ref{eq:out_p_const}) says that the resource sharing between \acp{CUE} and \acp{VUE} should ensure the coverage of \acp{VUE}. Here, $\gamma_0$ is the minimum SINR required for a reliable \ac{VUE} communication, and $p_0$ is its maximum outage probability.
    \item \textbf{\textit{Power Allocation Constraint}} - (\ref{eq:cell_power_const}) 
 and (\ref{eq:veh_power_const}) give the range of transmit powers of the \acp{CUE}  and \acp{VUE}, respectively, upper bounded by $P^\mathcal{C}_\mathrm{max}$ and $P^\mathcal{V}_\mathrm{max}$.
    \item \textbf{\textit{Resource Sharing Constraint}} - (\ref{eq:resource_sharing_const_1}) and (\ref{eq:resource_sharing_const_2}) ensure that the \acp{RB} of only one \ac{CUE} is shared with only one \ac{VUE}.
\end{itemize}}


\section{Proposed Power and Spectrum Allocation with Link Adaptation}\label{sec:Methodology}
 This section discusses the proposed resource allocation algorithms for the modes mentioned in Section \ref{sec:modeRB}, i.e.,  1) shared or underlay, and 2) dedicated or overlay. 
\subsection{\textbf{G}reedy \textbf{R}esource \textbf{A}llocation   \textbf{H}ungarian \textbf{S}haring }\label{sec:GRAHS}
{The first proposed algorithm, \textbf{G}reedy \textbf{R}esource \textbf{A}llocation with \textbf{H}ungarian \textbf{S}haring (\textbf{GRAHS}), which is outlined as Algorithm \ref{alg:GreedyAlloc} works as follows. (i) First, it executes a greedy approach to allocate resources to the \acp{CUE} according to their link condition (Section \ref{sec:GRA}). (ii)  Using this resource allocation, it allocates power to the \acp{CUE} and \acp{VUE} (Section \ref{sec:power_alloc}). (iii) With the new power allocation from step (ii), GRAHS pairs the \acp{CUE} and \acp{VUE}, using the maximum weighted bipartite matching Hungarian algorithm, such that each \ac{CUE}-\ac{VUE} pair meets the CUE rate constraint and the VUE coverage probability constraint. 
(iv)  Finally, GRAHS checks all the paired \acp{CUE} and \acp{VUE} to ascertain if the number of \acp{RB} required by the corresponding \ac{CUE} after pairing conforms to its respective QoS requirement or not. If not, it declares the pairing infeasible. The details follow.}
\begin{algorithm}[t]
\caption{Proposed \textbf{GRAHS}}\label{alg:GreedyAlloc}

\For{$t\in\{1,2,\cdots \mathrm{T_{obs}}\}$}{
Sort \ac{CUE} and \ac{VUE} packets into the buffers $\mathcal{B_C}$ and $\mathcal{B_V}$ according to their time-to-live.\;
\For {$c \in \mathcal{B_C} $}{
\uIf{ $\text{No. of \acp{CUE} scheduled}\leq C_t$\label{algoline:CUEcondcheck}}{
Find channel gain $g^n_{c,Z}=|h^n_{c,Z}|^2$ between the $c^\mathrm{th}$ \ac{CUE} and \ac{gNB} in \ac{RB}-$n$, $\forall n$ \label{algoline:calculateCSI}\;
{Calculate SNR $\gamma^{'n}_c = \frac{ P^{\mathit{C}}_c \alpha_{c,Z}|h^n_{c,Z}|^2}{\sigma^2}$  in each \ac{RB} and obtain the SE $R^{'n}_c$ from Table~\ref{tab:MCS} for a 
BLER of 0.1.}\label{algoline:calculaterateperRB}\;
 Sort all $N_1$ \acp{RB} in \ac{BWP}-1 in descending order of $R^{'n}_c$\;
{Find the greedy resource allocation ($\rho^{n}_c$)  that maximizes $
  R^{'}_c = \sum_{n=1}^{N_1} R^{'n}_c\rho^{n}_c$
 using minimum  possible number of \acp{RB}, $N_c$. \label{algoline:calculatenumRB} \label{algoline:findoptimalalloc}}\;
 Add $c$ to the scheduled \ac{CUE} list $\mathcal{S}_C$\label{algoline:CUEscheduled}\;
\For{$v \in \mathcal{B_V} $}{
For the pair $\{c,v\}$ obtain the optimal power allocation ${P^*_v, P^*_c}$ from (\ref{eq:optimalpowerveh}) and (\ref{eq:optimalpowercell})~\cite{Le2017}\label{algoline:poweralloc} \;
Obtain $R^*_{c}$ by substituting ${P^*_c, P^*_v}$\label{algoline:recalRc}\;
\If{$R^*_{c}<r_0$ \&\& $\mathtt{Pr}\{ \sum_n\gamma^n_v\rho^c_n x_{c,v} \leq \gamma_0\} > p_0$} {$R^*_{c} = -\inf$\ ;}
}
}
}
Use Hungarian algorithm to find optimal pairing $x_{c,v}$ between \acp{CUE} and \acp{VUE} based on $R^*_{c}$\label{algoline:resourcesharing}\;
\For{$c\in\mathcal{S_C}$}{
\For{$v\in\mathcal{B}_v$}{
\If{$x_{c,v}==1$}{
Calculate the SE per \ac{RB}, $ R^{n}_c $ and $R^{n}_v$ using the \ac{SINR} $\gamma^n_c$ and $\gamma^n_v$ from (\ref{eq:SINR_CUE}) and (\ref{eq:SINRVUE}), respectively\;
Calculate the number of \acp{RB} $ N^{*}_c $ and $ N^{*}_v $ using $ R^{n}_c $ and $R^{n}_v$\;
\If{$N^*_c = N_c$ \&  $N^*_v\leq N_c$} {Assign \acp{RB} with $\rho^{n}_c=1$ to \ac{VUE} v\;}
\Else{Scheduling the \ac{VUE} in next TTI\;}
}}}
\Return The  \ac{RB} allocation $\rho^{n^*}_c$ of \acp{CUE}, the optimal pairing $x_{c,v}$ between \acp{CUE} and \acp{VUE}, and the optimal power allocation ${P^*_c, P^*_v}$\; 
}
\end{algorithm}
\begin{table}[t]
\footnotesize
\caption{An example of achievable Spectral Efficiency of different users in different \acp{RB}. }\label{tab:SEexample}
\centering
\begin{tabular}{|p{0.5cm}|p{1.2cm}p{1.2cm}p{1.2cm}p{1.2cm}p{1.2cm}|}
\hline
\textbf{CUE Id} & \multicolumn{5}{c|}{\textbf{SE in each RB (bits/symbol)/ No. of bits per RB}}                                                                                  \\ \hline
        & \multicolumn{1}{p{1.2cm}|}{$\mathrm{RB_1}$}       & \multicolumn{1}{p{1.2cm}|}{$\mathrm{RB_2}$}         & \multicolumn{1}{p{1.2cm}|}{$\mathrm{RB_3}$}       & \multicolumn{1}{p{1.2cm}|}{$\mathrm{RB_4}$}       & {$\mathrm{RB_5}$}         \\ \hline
$\mathrm{c_1}$     & \multicolumn{1}{p{1.2cm}|}{1.18/198.24} & \multicolumn{1}{p{1.2cm}|}{1.18/198.24} & \multicolumn{1}{p{1.2cm}|}{1.48/248.64} & \multicolumn{1}{p{1.2cm}|}{1.48/248.64} & 1.91/320.88 \\ \hline
$\mathrm{c_2}$     & \multicolumn{1}{p{1.2cm}|}{1.91/320.88} & \multicolumn{1}{p{1.2cm}|}{1.91/320.88} & \multicolumn{1}{p{1.2cm}|}{1.91/320.88} & \multicolumn{1}{p{1.2cm}|}{1.48/248.64} & 1.48/248.64 \\ \hline
$\mathrm{c_3}$      & \multicolumn{1}{p{1.2cm}|}{1.48/248.64} & \multicolumn{1}{p{1.2cm}|}{1.48/248.64} & \multicolumn{1}{p{1.2cm}|}{1.48/248.64} & \multicolumn{1}{p{1.2cm}|}{1.48/248.64} & 1.18/198.24 \\ \hline
\end{tabular}
\end{table}
\subsubsection{\textbf{Resource Allocation  of \acp{CUE} }}\label{sec:GRA}
 \indent GRAHS runs at the beginning of every \ac{TTI}. \acp{CUE} and \acp{VUE} being real-time traffic users, their packets have expiry deadlines. So, GRAHS starts by sorting the \ac{CUE} and \ac{VUE} packets independently into two different buffers, $\mathcal{B_C}$ and $\mathcal{B_V}$, respectively, according to their time-to-live. This sends the earliest packet to the head of the queue, addressing the delay sensitivity.  It then allocates \acp{RB} to the \acp{CUE} only. First, GRAHS determines the channel gain $g^n_{c,Z}$ (\textit{Line \ref{algoline:calculateCSI}}) and calculates the \ac{SNR} of \ac{CUE} $c$ in \ac{RB}-$n$ as $\gamma^{'n}_c = \frac{ P^{\mathit{C}}_c g^n_{c,Z}}{\sigma^2} $, $\forall n$ (\textit{Line \ref{algoline:calculaterateperRB}}). At this point, GRAHS considers that a \ac{RB} is not yet shared with a \ac{VUE}. So, the \ac{SNR}, and not SINR, of the \acp{CUE}, is computed.  For the \ac{SNR} computed in \textit{Line - \ref{algoline:calculaterateperRB}}, GRAHS then uses Table \ref{tab:MCS} to find out the \ac{MCS} and the transmission rate $R^{'n}_c$ that it can use to serve the \ac{CUE} packet with a \ac{BLER} of 0.1. It uses this \ac{MCS} value to find the required number of \acp{RB}, $N_c$.\\
\indent {GRAHS then finds the greedy  resource allocation in \textit{Line-\ref{algoline:findoptimalalloc}}. To explain the greedy resource allocation, we use the example in Table \ref{tab:SEexample}}. Each row in the table represents a \ac{CUE} and its channel condition in the different \acp{RB}.  Each cell in Table \ref{tab:SEexample}, under the columns labeled \ac{RB}$_i$,  has an $x/y$ element, where `$x$' is the user's \ac{SE} in bits/symbol and `$y$' is the number of bits per \ac{RB}, for the \ac{SE} $x$. The \acp{CUE} in the leftmost column are sorted in the order of their shortest time-to-live., \textit{i.e.}, $c_1<c_2<c_3$. For each \ac{CUE}, we first sort the \acp{RB} in the descending order of achievable \acp{SE}. So, the sorted list for $c_1$ is $\mathrm{RB_5, RB_3, RB_4, RB_1, RB_2}$. The algorithm scans this sorted list and checks if the \ac{RB} with the highest \ac{SE} can accommodate the packet. This example shows that $\mathrm{RB_5}$ has the highest \ac{SE} of 1.91 bits/symbol for $c_1$. To serve a 50-byte packet with this \ac{SE}, two \acp{RB} will be needed. However, no other \ac{RB} has the same or higher \ac{SE}. Hence, the algorithm moves to the \ac{RB} with the second-highest \ac{SE}, which here is $\mathrm{RB_3}$. Its \ac{SE} is 1.48 bits/symbol.  It is found that for this \ac{SE} also, two \acp{RB} are needed to accommodate a \acp{CUE}'s packet. So, GRAHS assigns $\mathrm{RB_5}$, which has a higher \ac{SE} than $\mathrm{RB_3}$, and $\mathrm{RB_3}$ to $c_1$. In this case, the \ac{SE} used in both \acp{RB} is that of $\mathrm{RB_3}$, as $\mathrm{RB_3}$ will not be able to handle a higher \ac{SE}.   Subsequently, these \acp{RB} become unavailable to other \acp{CUE}. ~{If, however, the packet could not be accommodated within the two \acp{RB}, the search would have continued to identify three, four, or more \acp{RB} in which the packet could be served, albeit with a lower \ac{MCS}, according to Table \ref{tab:MCS}~\footnote{It may be noted that the explanation above also holds for $\mathrm{RB_5}$ and $\mathrm{RB_4}$, instead of $\mathrm{RB_5}$ and $\mathrm{RB_3}$ as $\mathrm{RB_4}$ and $\mathrm{RB_3}$ have the same \ac{SE}.}. In this way, GRAHS accommodates a \ac{CUE} packet using minimum number of \acp{RB} based on the first fit strategy to achieve a lower complexity. 
In such a multiple \ac{RB} assignment, the minimum achievable \ac{SE} among all assigned \acp{RB} is used. This is because a \ac{RB} which has a higher \ac{SNR} can support the lower \ac{SE}, but the reverse is not true~\cite{Palit2015}.} \\
\indent This part of the algorithm returns the RB allocation variable $\rho^{n}_c, \forall n$, of the \acp{CUE}. The \acp{CUE} which are allocated \acp{RB} in  \textit{Line-\ref{algoline:findoptimalalloc}}, are added to the scheduled list $\mathcal{S}_c$ in \textit{Line-\ref{algoline:CUEscheduled}}. {An important point to note is that the maximum number of users that can be scheduled in a \ac{TTI} is limited by the availability of the \ac{PUCCH}~\cite{TS36213}. So, we have assumed that the number of users that can be scheduled at a time is limited by $C_t$, (Line-\ref{algoline:CUEcondcheck}). The algorithm next pairs the \acp{CUE} and \acp{VUE} for \ac{RB} sharing.} 
\subsubsection{\textbf{Resource Sharing between \acp{CUE} and \acp{VUE}}}: The resource sharing between \acp{CUE} and \acp{VUE} includes: a) the power allocation to the \ac{CUE}-\ac{VUE} pairs so that the interference is limited (\textit{Line - \ref{algoline:poweralloc}}), and b) selection of the optimal \ac{CUE}-\ac{VUE} pairs (\textit{Line - \ref{algoline:resourcesharing}}). 
\paragraph{\textbf{Optimal Power Allocation}}\label{sec:power_alloc} {After \ac{RB} allocation, GRAHS allocates power to all the \ac{CUE}-\ac{VUE} pairs in the \acp{RB} assigned to the \acp{CUE}~\cite{Le2017}. In other words, for the \ac{CUE}-$c$, GRAHS  allocates power to all the $V$ \ac{VUE} connections in the \acp{RB} assigned to $c$, i.e., those with $\rho_c^n=1$. The power is allocated in a way that maximizes the \ac{CUE} rate while satisfying the coverage probability constraint of \acp{VUE} in (\ref{eq:out_p_const})~\cite{Le2017}. As mentioned earlier, after \ac{RB} allocation, the minimum achievable transmission rate among all the \acp{RB} assigned to a \ac{CUE} is used in each \ac{RB}. So, the objective of the power allocation is to maximize this minimum transmission rate. We denote the minimum transmission rate as  $R_{c,\eta}$ 
\begin{equation}
    R_{c,\eta}=  B \log_2(1+\frac{\po{C}{c}\alpha_{c,Z}|h^{\eta}_{c,Z}|^2}{\sigma^2+\po{V}{v}\tilde{\alpha_v}|\Tilde{h}^{\eta}_v|^2}),
\end{equation} 
where $ \eta=\underset{n}{\mathrm{argmin}} (R^n_{c}),\forall n\ \text{with}\ \rho^n_c=1$. Thus, the problem can be rewritten in this part as:
\begin{align}
&(P^*_c,P^*_v)= \underset{\po{C}{c},\po{V}{v}}    {\mathrm{argmax}}\ R_{c,\eta}\label{eq:opt_power_alloc}\\
\text{s. t. \ } &(\ref{eq:out_p_const}),   \text{(\ref{eq:cell_power_const})\ and (\ref{eq:veh_power_const})) are satisfied.}\nonumber
\end{align}
}
The optimal power allocation from (\ref{eq:opt_power_alloc}) is: 
\begin{align}\label{eq:optimalpowerveh}
    P^*_v = 
    \begin{cases}
        \mathrm{min}\{P^\mathcal{V}_\mathrm{max},P^{\mathcal{V}_1}_{\mathcal{C},\mathrm{max}}\}, \ \text{if} \ \po{V}{\mathrm{max}}\leq\po{V}{0},\\
        \mathrm{min}\{P^\mathcal{V}_\mathrm{max},P^{\mathcal{V}_2}_{\mathcal{C},\mathrm{max}}\}, \ \text{if} \ \po{V}{\mathrm{max}}>\po{V}{0}, \text{and} \ \po{C}{\mathrm{max}}>\po{C}{0},\\
P^{\mathcal{V}_1}_{\mathcal{C},\mathrm{max}}, \ \text{otherwise},
    \end{cases}
\end{align}
and
\begin{align}\label{eq:optimalpowercell}
    P^*_c = 
    \begin{cases}
     \mathrm{min}\{\po{C}{\mathrm{max}},P^{\mathcal{C}_1}_{\mathcal{V},\mathrm{max}}\}, \ \text{if} \ \po{V}{\mathrm{max}}\leq\po{V}{0},\\  
      \mathrm{min}\{\po{C}{\mathrm{max}},P^{\mathcal{C}_2}_{\mathcal{V},\mathrm{max}}\}, \ \text{if} \ \po{V}{\mathrm{max}}>\po{V}{0}, \text{and} \ \po{C}{\mathrm{max}}>\po{C}{0},\\ 
      \po{C}{\mathrm{max}}, \ \mathrm{otherwise},
    \end{cases}
\end{align}
where
\begin{equation}
    \po{C}{0} = \frac{\sigma^2}{\frac{1-\epsilon^2_{c,v}}{1-\epsilon^2_v}\left(\frac{1}{p_0}-1\right)\alpha_{c,v}\epsilon^2_v|\hat{h}^{\eta}_v|^2-\alpha_{c,v}\epsilon^2_{c,v}|\hat{h}^{\eta}_{c,v}|^2}
\end{equation}
and
\begin{equation}
    \po{V}{0} = \frac{\po{C}{0}\gamma_0\alpha_{c,v}(1-\epsilon^2_{c,v})(1-p_0)}{\alpha_v(1-\epsilon^2_v)p_0}
\end{equation}
$P^{\mathcal{V}_1}_{\mathcal{C},\mathrm{max}}$ and $P^{\mathcal{C}_1}_{\mathcal{V},\mathrm{max}}$ are derived from the functions,
\begin{equation}
\mathcal{F}_1(P^{\mathcal{V}_1}_{\mathcal{C},\mathrm{max}},\po{C}{\mathrm{max}}) = 0 \ \text{and} \ \mathcal{F}_1(\po{V}{\mathrm{max}}, P^{\mathcal{C}_1}_{\mathcal{V},\mathrm{max}}) = 0.
\end{equation}
$P^{\mathcal{V}_2}_{\mathcal{C},\mathrm{max}}$ and $P^{\mathcal{C}_2}_{\mathcal{V},\mathrm{max}}$ are derived from the functions,
\begin{equation}
\mathcal{F}_2(P^{\mathcal{V}_2}_{\mathcal{C},\mathrm{max}},\po{C}{\mathrm{max}}) = 0 \ \text{and} \ \mathcal{F}_2(\po{V}{\mathrm{max}}, P^{\mathcal{C}_2}_{\mathcal{V},\mathrm{max}}) = 0.
\end{equation}
through bisection search by noting the monotonic relation between $\po{C}{c}$, $\po{V}{m}$ in the functions,
\begin{equation*}
    \mathcal{F}_1\left(\po{C}{c},\po{V}{v}\right)=\exp\left(\frac{G\gamma_0}{F}\right)\left(1+\frac{H}{F}\gamma_0\right)-\frac{\exp(\frac{A}{F})}{1-p_0}=0
\end{equation*}
when $\po{V}{v}\in(0,\po{V}{0})$ and 
\begin{equation*}
 \mathcal{F}_2\left(\po{C}{c},\po{V}{v}\right)   = \left(1+\frac{F}{\gamma_0 H}\right)\exp\left(\frac{A-G\gamma_0}{\gamma_0 H}\right)-\frac{1}{p_0}=0
\end{equation*}
when $\po{V}{v}\in(\po{V}{0},+\infty)$. Here, 
$A=\po{V}{v} \alpha_v \epsilon^2_v|\hat{h}^{\eta}_v|^2$,
$F=\po{V}{v}\alpha_v(1-\epsilon^2_v)$,
$G=\sigma^2+P^{\mathit{C}}_c\alpha_{c,v}\epsilon^2_{c,v}|\hat{h}^{\eta}_{c,v}|^2$ and $H=P^{\mathit{C}}   _c\alpha_{c,v}(1-\epsilon^2_{c,v})$.
\paragraph{\textbf{Hungarian algorithm for Resource Sharing (HS)}}\label{sec:hungsharing}
{With the new power allocation ($P^*_c,P^*_v$), the minimum transmission rate of the \acp{RB} assigned to a \ac{CUE} is recalculated as  $R^*_{c,\eta}=  B \log_2(1+\frac{P^*_c\alpha_{c,Z}|h^{\eta}_{c,Z}|^2}{\sigma^2+P^*_v\tilde{\alpha_v}|\Tilde{h}^{\eta}_v|^2})$. Consequently, GRAHS obtains  the new \ac{CUE} rate $R_c^*=R^*_{c,\eta}\sum_n \rho^n_c$ (\textit{Line \ref{algoline:recalRc}}). The \acp{VUE} for which $R_c^*$ meets the rate constraint of (\ref{eq:rate_const}) and which meet the coverage probability constraint of (\ref{eq:out_p_const}) are added to the candidate \ac{VUE} list for resource sharing with  \ac{CUE} $c$.}  After all the \acp{CUE} and \acp{VUE} are scanned, they need to be paired in such a way that maximizes the transmission rate of the \acp{CUE}. This pairing problem is one of maximum weighted bipartite matching in which the two sets are the \acp{CUE} and the \acp{VUE}. The weights of the links between these sets are the transmission rates of the \acp{CUE}. The maximum weight perfect matching of these sets can be obtained in polynomial time using the Hungarian algorithm~\cite{Le2017,West2000}, which we have done in this work. This step returns the pairing variable {$x_{c,v}, \forall c, \forall v$}. 
\subsubsection{\textbf{Final Allocation}}\label{sec:finalalloc}
\indent At this stage, GRAHS has the \ac{RB} allocation of the \ac{CUE}-$c$ stored in $\rho^n_c$ and the pairing of the \ac{CUE}-$c$ and \ac{VUE}-$v$ stored in $x_{c,v}$. It next checks whether the pairs are viable with respect to their \ac{RB} requirement.\\
\indent GRAHS picks each $\{c,v\}$ pair with {$x_{c,v}=1$} and recalculates the \acp{SINR} $\gamma^n_c$ and $\gamma^n_v$ from (\ref{eq:SINR_CUE}) and  (\ref{eq:SINRVUE}), respectively, in the \acp{RB} which are assigned to \ac{CUE}-$c$ only (\textit{i.e.}, the \acp{RB} with $\rho^n_c=1$). 
Thus, at this point, the algorithm accounts for the interference between the \acp{CUE} and \acp{VUE}. 
It then uses Table \ref{tab:MCS} to calculate the number of \acp{RB}, $N^*_c$,  needed to serve a \ac{CUE} packet with a BLER of 0.1 using the newly calculated \ac{SINR} $\gamma^n_c$  (\ref{eq:SINR_CUE}). Similarly, it also estimates  $N^*_v$, \textit{i.e.}, the number of \acp{RB} needed to serve a \ac{VUE} packet with a BLER of 0.001 using the newly calculated \ac{SINR} $\gamma^n_v$  (\ref{eq:SINRVUE}). To find $N^*_c$ and $N^*_v$, it uses the same method as explained in Section \ref{sec:LA}. 
The \ac{CUE}-\ac{VUE} pair $\left(c,v\right)$ is finalized if and only if $N_c^* = N_c$ and $N_v^*\leq N_c$. Otherwise, the algorithm assigns resources to the CUE only and defers the allocation of the VUE till the next slot. This step returns the updated sharing variable $x^*_{c,v}$.


\subsection{\textbf{H}ungarian \textbf{R}esource \textbf{A}llocation \textbf{H}ungarian \textbf{S}haring }\label{sec:HRAHS}
A major drawback of GRAHS is that it does not allocate contiguous \acp{RB} to the users.  However, 3GPP advocates contiguous resource block allocation, especially in the Frequency Range 2~\cite{TS38101,TS38214,Dahlamn2016}. So, we have designed a second resource allocation algorithm for the \acp{CUE}, in which instead of treating each \ac{RB} independently, we have grouped a few \acp{RB} into one Resource Chunk (\ac{RC})~\cite{Calabrese2008,Mukho2015}. The algorithm assigns one \ac{RC} to one \ac{CUE} in each \ac{TTI} such that the sum rate of the \acp{CUE} is maximized. Thus, this assignment problem of the \acp{CUE} to \acp{RC} is also one of maximum weight bipartite matching, in which the sets to be matched are the \acp{CUE} and the \acp{RC}, and the weights are the rates of the \ac{CUE}. So, we have used the Hungarian algorithm for assigning \acp{RC} to \acp{CUE}. Therefore, the second algorithm of our work uses the Hungarian assignment in two stages - first to assign  \acp{RC} to the \acp{CUE}, and then to pair the \acp{CUE} with the \acp{VUE}. We call this algorithm \textit{\textbf{HRAHS}} - Hungarian Resource Allocation with Hungarian Sharing, which is outlined as Algorithm \ref{alg:HRAHS}.\\
\indent We have observed from extensive simulations of GRAHS (10 simulation runs, each of 6.25 seconds)  that on an average 2.14, \textit{i.e.}, three \acp{RB} are required to serve a packet of size 50 bytes for a BLER of 0.1. Furthermore, we have also observed that nearly 92.78\% of users needed fewer than four \acp{RB}. Hence, we have grouped four \acp{RB} to form a \ac{RC}. It is to be noted that the size of the \acp{RC} is a design choice and may be changed based on the network scenario.\\
\begin{algorithm}[t]
\caption{Proposed \textbf{HRAHS}}\label{alg:HRAHS}

\For{$t\in\{1,2,\cdots T\}$}{
Sort \ac{CUE} and \ac{VUE} packets into the buffers $\mathcal{B_C}$ and $\mathcal{B_V}$ according to their time-to-live.\;
\If{$\mathtt{len}(\mathcal{B_C})<C_t$\label{algo2Line:shortBuffer}}{Append $\{C_t-\mathtt{len}(\mathcal{B_C})\}$ null users to $B_C$\;}
\For {$c \in \mathcal{B_C}[1:C_t]$\label{algo2Line:takeCt}}{
Find the \ac{CSI} $g^j_{c,Z}$ in \ac{RC}-$j$, $\forall j$ \acp{RC}, between the $c^\mathrm{th}$ \ac{CUE} and the \ac{gNB}\label{algo2line:calculateCSI}\;
Using $g^j_{c,Z}$, calculate the rate  $
  R^j_c = \log_2(1+\frac{ \po{C}{c} \alpha_{c,Z}|h^j_{c,Z}|^2}{\sigma^2})
 $ in each \ac{RC}.\label{algo2line:calculaterateperRC}\;}
 Use a maximizing Hungarian algorithm to find the optimal pairing between $C_t$ \acp{CUE} of $\mathcal{B_C}$ and the $C_t$ \acp{RC} so as to maximize
  $\sum_j\sum_c R^j_c$.\label{algo2line:findoptimalalloc} while providing a \ac{BLER}=0.1\;
  Add scheduled \acp{CUE} to the list $\mathcal{S_C}$\;
  \For{$c\in S_c$}{
\For{$v \in \mathcal{V} = \{1,2,\cdots,V\} $}{
For the pair  $\{c,v\}$ obtain the optimal power allocation ${P^*_v, P^*_c}$ from (\ref{eq:optimalpowerveh}) and (\ref{eq:optimalpowercell})~\cite{Le2017}\label{algo2line:poweralloc}\;
Obtain $R^*_{c}$ by substituting ${P^*_c, P^*_v}$\label{algo2line:recalRc}\;
\uIf{$R^*_{c}<r_0$} {$R^*_{c} = -\inf$\;}
}}
Use Hungarian algorithm to find optimal pairing $x_{c,v}$ between \acp{CUE} and \acp{VUE} based on $R^*_{c}$\label{algo2Line:Hungarianforpairing}\;
\For{$c\in\mathcal{S_C}$}{
\For{$v\in\mathcal{B}_v$}{
\If{$x_{c,v}==1$}{
Calculate the SE per {\ac{RB}}, $R^{n}_c$ and $R^{n}_v$, for the \textbf{\acp{RB}} of the \ac{RC} allocated to \ac{CUE} $c$ ($\zeta^j_c=1$) using the \ac{SINR} $\gamma^n_c$ and $\gamma^n_v$ from (\ref{eq:SINR_CUE}) and (\ref{eq:SINRVUE}), respectively\;
Calculate the number of \acp{RB}  $N^{*}_v $ using $ R^{n}_c $ and $R^{n}_v$ for BLER = 0.001\;
\If{$N^*_v\leq \text{size(RC)}$} {Assign $N^*_v$ \acp{RB}, from the \ac{RC} allocated to \ac{CUE} $c$, to the \ac{VUE} v\;}
\Else{Scheduling of \ac{VUE} is deferred to the next slot\;}
}}}
\Return The optimal \ac{RC} allocation $\zeta^{j*}_c$ of \acp{CUE}, the optimal pairing $x^*_{c,v}$ between \acp{CUE} and \acp{VUE}, the optimal power allocation ${P^*_c, P^*_v}$, and the \ac{RB} allocation of \acp{VUE}\; 
}
 \end{algorithm}
\indent HRAHS runs at the beginning of a TTI and sorts the \ac{CUE} and \ac{VUE} packets according to their time-to-live. into the two independent buffers, $\mathcal{B_C}$ and $\mathcal{B_V}$, respectively. Once the packets are sorted, it first allocates resources to the \acp{CUE} without considering any interference from the \acp{VUE}. We represent the \ac{SNR} (\ac{SINR}) of \ac{CUE} $c$ in \ac{RC} $j$ as $\gamma^{j'}_c$ ($\gamma^{j}_c$). We introduce a new \ac{RC} allocation variable $\zeta^j_c$, such that $ \zeta^j_c= 1, \text{if \ac{CUE} \textit{c} is allocated to \ac{RC} \textit{j} }$, and $\zeta^j_c =  0$, otherwise, $\forall c\in \mathcal{C}$ and $\forall j \in \mathcal{J}$. Here, $\mathcal{J}$ gives the set of all \acp{RC}. We assume, the number of \acp{RC} in $\mathcal{J}$ is equal to the maximum number of users that can be scheduled in a \ac{TTI}, which here is $C_t$. Thus, the resource allocation problem of \acp{CUE} reduce to:
\begin{equation}
     \max_{\{\zeta^j_c\}} \sum_{j\in\mathcal{J}}\sum_{c\in\mathcal{C}} R^j_c =  \log_2(1+\gamma^{j'}_c).
     \label{eq:Hungarian_RA_CUE}
\end{equation}
(\ref{eq:Hungarian_RA_CUE}) is, thus, a $C_t\times C_t$ maximum weight bipartite matching problem, which matches $C_t$ users  to $C_t$ \acp{RC} with the objective of maximizing $R_c$. In some \acp{TTI}, however, the number of \acp{CUE} in the buffer $\mathcal{B_C}$ can be less or more than $C_t$. If it is less than $C_t$, then HRAHS first appends $C_t-|\mathcal{B_C}|$ null users to $\mathcal{B_C}$ (\textit{{Line-\ref{algo2Line:shortBuffer} Algorithm 2}}) in order to execute the Hungarian algorithm. Otherwise, it selects the first $C_t$ users for allocation to the \acp{RC} (\textit{{Line-\ref{algo2Line:takeCt} Algorithm 2}}).\\
\indent HRAHS next allocates power to each \ac{CUE}-\ac{VUE} pair in the \acp{RC} (Line-\ref{algo2line:poweralloc}) so as to maximize the \ac{CUE} datarate, as in GRAHS. It then recalculates the \ac{CUE} rate $R^*_c$ with the new power allocation, and uses $R^*_c$  to pair the \acp{CUE} and \acp{VUE} for sharing the \acp{RC}, using Hungarian algorithm (\textit{Line-\ref{algo2Line:Hungarianforpairing} for Algorithm 2}). The pairing using Hungarian algorithm ensures that the \ac{CUE} rate is maximized while satisfying both the \ac{CUE} rate constraint and the \ac{VUE} coverage probability constraints. generating the pairing variable $x_{c,v}$. \\
\indent Once the pairing is done,  the algorithm then selects each $\{c,v\}$ pair and recalculates the \ac{SINR}, $\gamma^n_c$ and $\gamma^n_v$,  in all the \acp{RB} of the \ac{RC} assigned to \ac{CUE} $c$, (\textit{i.e.}, the \acp{RB} in the \ac{RC} having $\zeta^j_c=1$). It then uses $\gamma^n_v$  to calculate the number of \acp{RB}, $N^*_v$, needed to transmit a \ac{VUE} packet at a BLER = 0.001.  If $N^*_v$ is less than or equal to the size of a \ac{RC}, the \ac{CUE} $c$ and \ac{VUE} $v$ are paired. Otherwise, the \ac{VUE} $v$ is scheduled in the next \ac{TTI}. If $N^*_v$ is less than the size of a \ac{RC}, then $N^*_v$ \acp{RB} from the \ac{RC} allocated to \ac{CUE} $c$ are assigned to the \ac{VUE} $v$. This method returns the optimal \ac{RC} allocation $\zeta^{j*}_c$, the optimal pairing $x^*_{c,v}$, and the optimal power allocation $P^*_c, P^*_v$.
\subsection{\textbf{O}verlay \textbf{R}esource \textbf{A}llocation (ORA)}\label{sec:ORA}
To establish the performance of GRAHS and HRAHS, we have compared them to the dedicated or the \textbf{O}verlay mode of \textbf{R}esource \textbf{A}llocation -- ORA. In ORA, \acp{CUE} and \acp{VUE} do not share resources. Instead, \acp{VUE} are also assigned dedicated \acp{RB}. So, \acp{CUE} and \acp{VUE} packets are first sorted into a single array according to their time-to-live.. Then, their \acp{SNR} (no resource sharing, hence, no interference) are obtained in all the \acp{RB}, and the number of \acp{RB} needed to serve a \ac{CUE} and a \ac{VUE} packet is calculated as in Section \ref{sec:LA}. Finally, similar to GRAHS, a greedy resource allocation approach is adopted to assign the \acp{RB} to the users. As \acp{VUE} are also assigned dedicated \acp{RB}, hence the limit on the number of users that can be scheduled in a TTI applies to both \acp{CUE} and \acp{VUE}.
\subsection{Resource Allocation of Best Effort Users}
For the resource allocation algorithms - GRAHS (Section \ref{sec:GRAHS}), HRAHS (Section \ref{sec:HRAHS}), ORA (Section \ref{sec:ORA})- a separate \ac{BWP} is reserved for the \acp{BUE}.  This \ac{BWP}-2, as mentioned in Section \ref{sec:sys_model}, has a SCS of 15 KHz and a time slot of 1 ms. So, the bandwidth of each \ac{RB} is 180 KHz. These \acp{RB} are allocated to the \acp{BUE} using Max C$\backslash$I algorithm \cite{dahlman2013}. In a \ac{TTI}, we assign each \ac{RB} to that \ac{BUE} which has the highest \ac{CSI} in that \ac{RB}. The resultant scheme is not fair, but it maximizes the sum rate of the \acp{BUE}. 
\subsection{Complexity Analysis}
We next compare the algorithms in terms of their time complexity. \\
For each \ac{CUE}, GRAHS executes the following steps.
\begin{enumerate}
    \item It sorts the \acp{RB} according to the \acp{SE} achievable by the \ac{CUE}. When merge-sort is used, the worst-case complexity of this sorting is $\mathcal{O}(N\log N)$.
    \item It then calculates the number of \acp{RB} required to serve the packet and identifies these \acp{RB}. According to Table \ref{tab:MCS}, the maximum number of \acp{RB} required to serve a \ac{CUE} packet of 50 bytes with a BLER of 0.1 is 16. So, in the worst case, GRAHS scans the first 16 \acp{RB} in the sorted list of \acp{RB}.
    \item It checks all the \acp{VUE} to find the optimal pairing.
\end{enumerate}
\indent   As a maximum of $C_t$ users can be scheduled in a \ac{TTI}, the worst case time complexity for \ac{CUE} resource allocation  and user pairing of GRAHS is $\mathcal{O}(C_t(N_1\log N_1+16+V))$. The Hungarian resource-sharing algorithm in GRAHS matches $C_t$ \acp{CUE} to $C_t$ \acp{VUE}. So, the complexity of Hungarian resource sharing is $\mathcal{O}(C^3_t)$. After pairing, the number of \acp{RB} needed individually by the \ac{CUE} and the \ac{VUE} packet of the pair is recalculated. Therefore, the complexity of all steps of GRAHS is $\mathcal{O}(C_t(N_1\log N_1+16+V)+C^3_t+2C_t)$, \textit{i.e.}, $\mathcal{O}(C_t^3)$\\
\indent In HRAHS, the resource allocation of \acp{CUE} is  carried out using the Hungarian algorithm. Since a maximum of $C_t$ users is assigned to $C_t$ \acp{RC}, hence, the complexity of the \ac{CUE} resource allocation algorithm in HRAHS is $\mathcal{O}(C^3_t)$. Again, $C_t$ \acp{VUE} are paired with $C_t$ \acp{CUE} using the Hungarian algorithm, and after pairing, the number of \acp{RB} needed by the \ac{VUE} of each pair is recalculated. So, the combined complexity is $\mathcal{O}(2C^3_t+V+2C_t)$, \textit{i.e.}, $\mathcal{O}(C^3_t)$
\section{Evaluation}\label{sec:Results}
In this section, we first explain the simulation parameters and then discuss the results obtained. 
\subsection{Simulation Parameters}
\begin{table}[t]
\centering
\caption{Simulation Parameters}\label{tab:sim_param}
\begin{tabular}{|c | c |}
\hline
\textbf{Parameter} & \textbf{Value}  \\
\hline  
Carrier Frequency & 28GHz  \\
\hline
\ac{gNB} / Vehicle  antenna gain & 8 dBi / 3 dBi \\
\hline
Simulation Area / Lane width  & 1 Km$\times$1 Km / 4 m \\
\hline
 Number of Lanes & 4 lanes in each direction \\
\hline
BS / Vehicle Noise Figure &  5 dB /  9 dB  \\
\hline
Noise Power $ \sigma^2$  &  -114 dBm \\
\hline
Vehicle speed &  50 Kmph \\
\hline
Minimum spectral efficiency of \acp{CUE} $r_0$ &  0.5 bps/Hz \\
\hline
SINR Threshold for coverage of \acp{VUE} $\gamma_0$ &  5 dB \\
\hline
Outage probability  $p_0$ &  $10^{-3}$ \\
\hline
Maximum \acp{CUE}/\acp{VUE} transmit power $P_c$,$P_v$  &  23 dBm \\
\hline
Shadowing type &  Log-normal \\
\hline
Shadowing standard deviation V2V / V2I & 4 dB / 7.8 dB\\
\hline
Number of \ac{VUE} Pairs / \acp{BUE} & 10 / 10 \\
\hline

Delay constraint of \acp{VUE} ($\delta_v$)/CUE ($\delta_c$) & 10 msec / 50 msec \\
\hline
Packet size CUE ($\beta_c$) / \acp{VUE}  ($\beta_v$) & 50 bytes / 10 bytes \\
\hline
Packet generation rate of CUEs $\lambda_c$  & Number of \acp{CUE}/20 \\
\hline
Feedback Period $T$ & 0.125 ms\\
\hline
\end{tabular}
\end{table}
The performances of the GRAHS and HRAHS algorithms have been validated using extensive Monte Carlo Simulations in a system-level simulator developed using Python. Each point in the figures corresponds to an average of 10 different simulation runs, each of which is 6.25 seconds long. {The simulation scenario considers the service area of a single \ac{gNB} to be a $\mathrm{1km\times 1km}$ square area with the \ac{gNB} at the center. We have considered eight lanes, each of width $w=4$ metres (m) and located 35m to the south of the \ac{gNB}.} All simulation parameters have been tabulated in Table \ref{tab:sim_param}. Some of the more important ones are discussed here.
\indent We have considered a high frequency of operation - 28 GHz.  As mentioned before, the \ac{QoS} constrained users (\acp{CUE}/\acp{CUE}+\acp{VUE}) are assigned to \ac{BWP}-1 which uses numerology $\mu=3$. The maximum number of  these users scheduled in a \ac{TTI} is considered to be eight. So, a minimum of eight \acp{RC} will be needed in HRAHS. Accordingly, we have taken the bandwidth of \ac{BWP}-1 to be -- \textit{number of RCs}$\times$\textit{ number of RBs in each RC }$\times$\textit{ bandwidth of each RB} = $8\times4\times1440 = 46.08$ MHZ. \ac{BWP}-2 has a bandwidth of 3.92 MHz, a carrier frequency of 2GHz, and uses numerology $\mu=0$.  The pathloss model used for \acp{CUE} is  $\mathrm{32.4 + 20\log_{10}(f_c) + 30log_{10}(d)}$  and for \acp{VUE} is $\mathrm{36.85 + 30 \log_{10}(d) + 18.9  \log_{10}(f_c)}$~\cite{TS37885}.

We have analyzed the performance of the proposed algorithms based on the real-time traffic capacity, which depends on the number of satisfied  real-time traffic users. Real-time traffic, such as VoIP calling, video calling, video streaming, etc., are characterized by their \ac{QoS} requirements, such as delay deadlines, maximum allowable packet loss rate, etc.  A real-time traffic user is satisfied if  (a) at least 98\% of its packets are successfully delivered, \textit{i.e.}, its packet loss ratio is less than 2\%, and (b) the average packet delay is less than the delay deadline. The real-time traffic capacity is given by that number of users, out of which 95\% are satisfied~\cite{Palit2015}. In this work, we have found the real-time traffic capacity for the given 5G network in terms of the number of satisfied \acp{CUE} while keeping the number of \acp{BUE} and \ac{VUE} pairs constant. We have considered ten \acp{BUE} and ten \ac{VUE} pairs, \textit{i.e.}, ten VUE transmitters and ten VUE receivers. 
\subsection{Results}
\indent In this section, we have first established the need for link-adaptation by comparing the performance of GRAHS, HRAHS, and ORA with an existing algorithm~\cite{Le2017} which assigns one \ac{RB} per \ac{CUE}. Subsequently, we have compared the performance of the GRAHS, HRAHS, and ORA algorithms in terms of their real-time traffic capacity, sum-rate of \acp{CUE}, and \ac{QoS} performance of \acp{CUE} and \acp{VUE}. 
\subsubsection{\textbf{Need for Link Adaptation}}
\begin{figure}[h]
\centering
\begin{subfigure}{0.25\textwidth}
         \centering
         \includegraphics[trim={0.5cm 1cm 0cm 2cm},width=\textwidth]{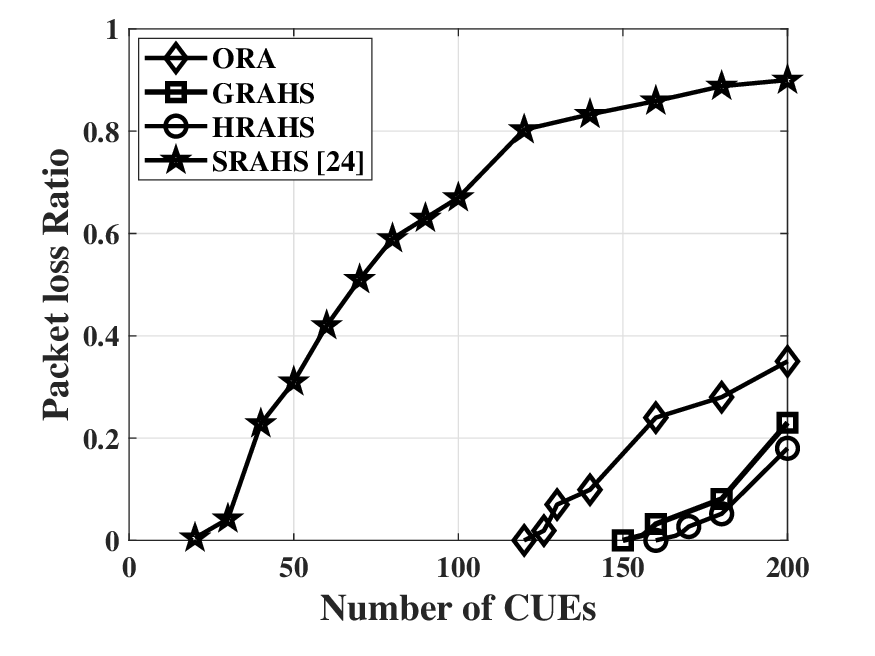}
\caption{Packet Loss Rate of \acp{CUE} }
\label{fig:LAutility_PLR}
     \end{subfigure}
 \hspace*{-0.5cm}       
     \begin{subfigure}{0.24\textwidth}
         \centering
         \includegraphics[trim={0cm 0cm 1cm 1.25cm},width=\textwidth]{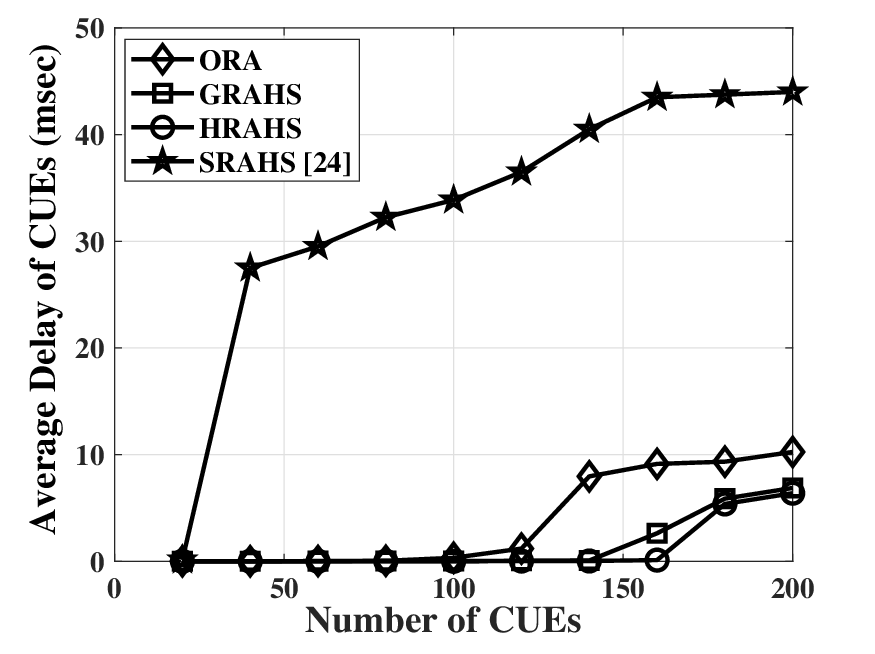}
\caption{Average Delay of \acp{CUE} }
\label{fig:LAutility_delay}
     \end{subfigure}
        \caption{Comparison of ORA, GRAHS, HRAHS with baseline SRAHS~\cite{Le2017} that has no link adaptation}
        \label{fig:CUE_QoS_LA}
\end{figure}
\begin{figure*}[t]
     \centering
          \begin{subfigure}{0.31\textwidth}
         \centering
         \includegraphics[trim={0cm 0cm 0cm 0cm},width=0.995\textwidth]{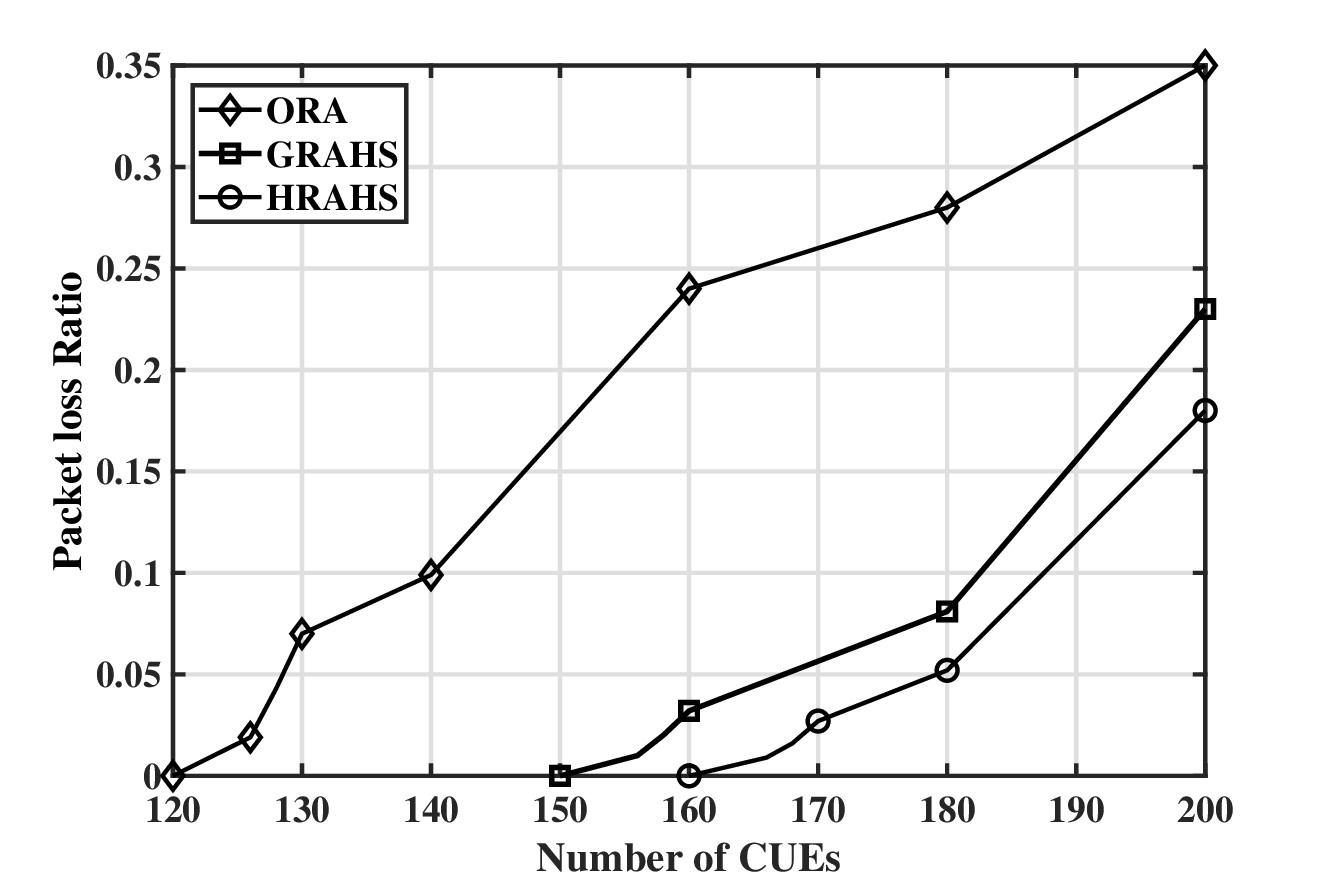}
         \caption{Packet Loss Ratio vs number of \acp{CUE}}
         \label{fig:CUE_PLR}
     \end{subfigure}
     \hspace*{-0.4cm}
     \begin{subfigure}{0.31\textwidth}
         \centering
         \includegraphics[width=1\textwidth]{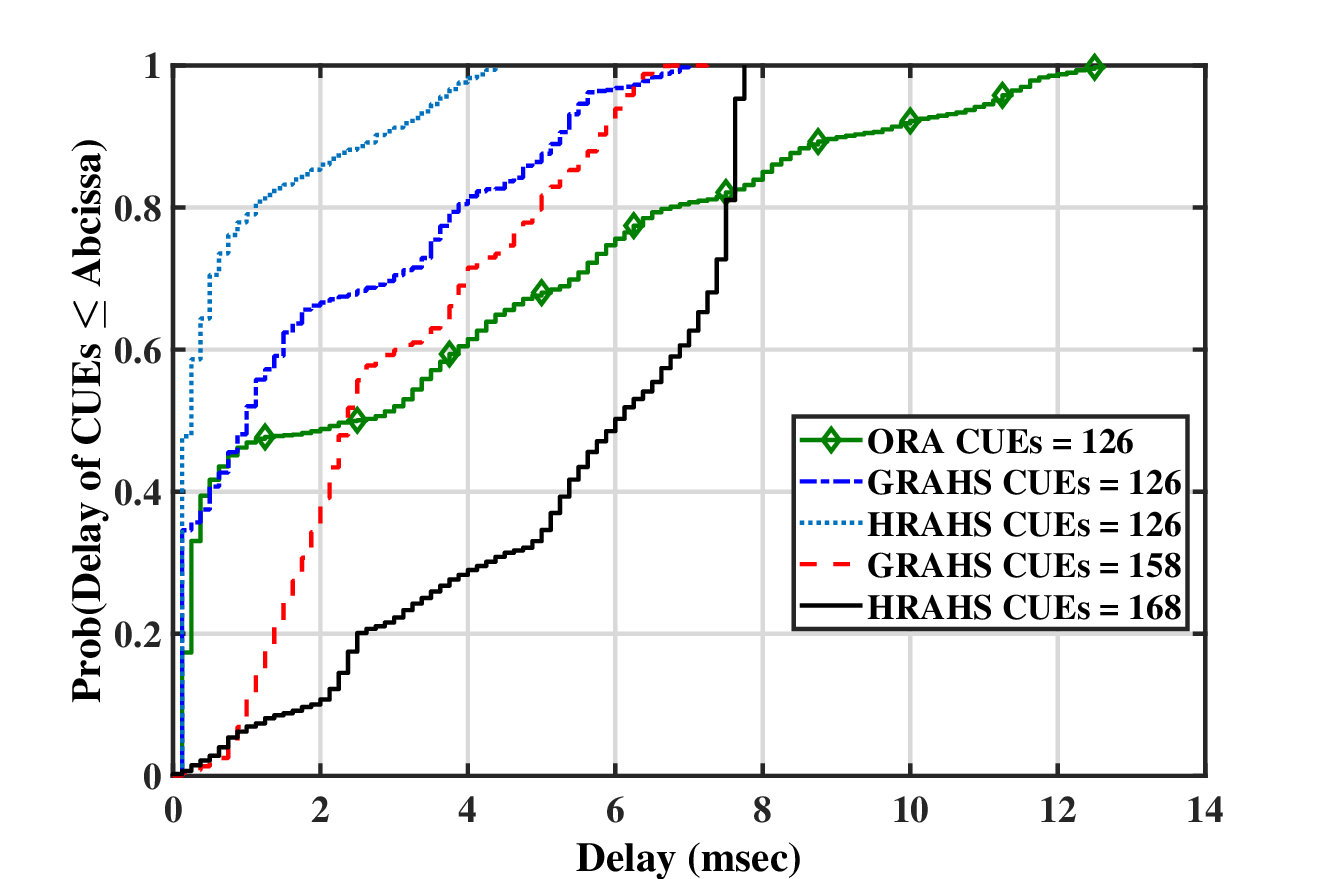}
         \caption{CDF of Packet Delay of \acp{CUE}}
         \label{fig:CUE_delay}
     \end{subfigure}
 \hspace*{-0.4cm}       
     \begin{subfigure}{0.31\textwidth}
         \centering
         \includegraphics[width=1\textwidth]{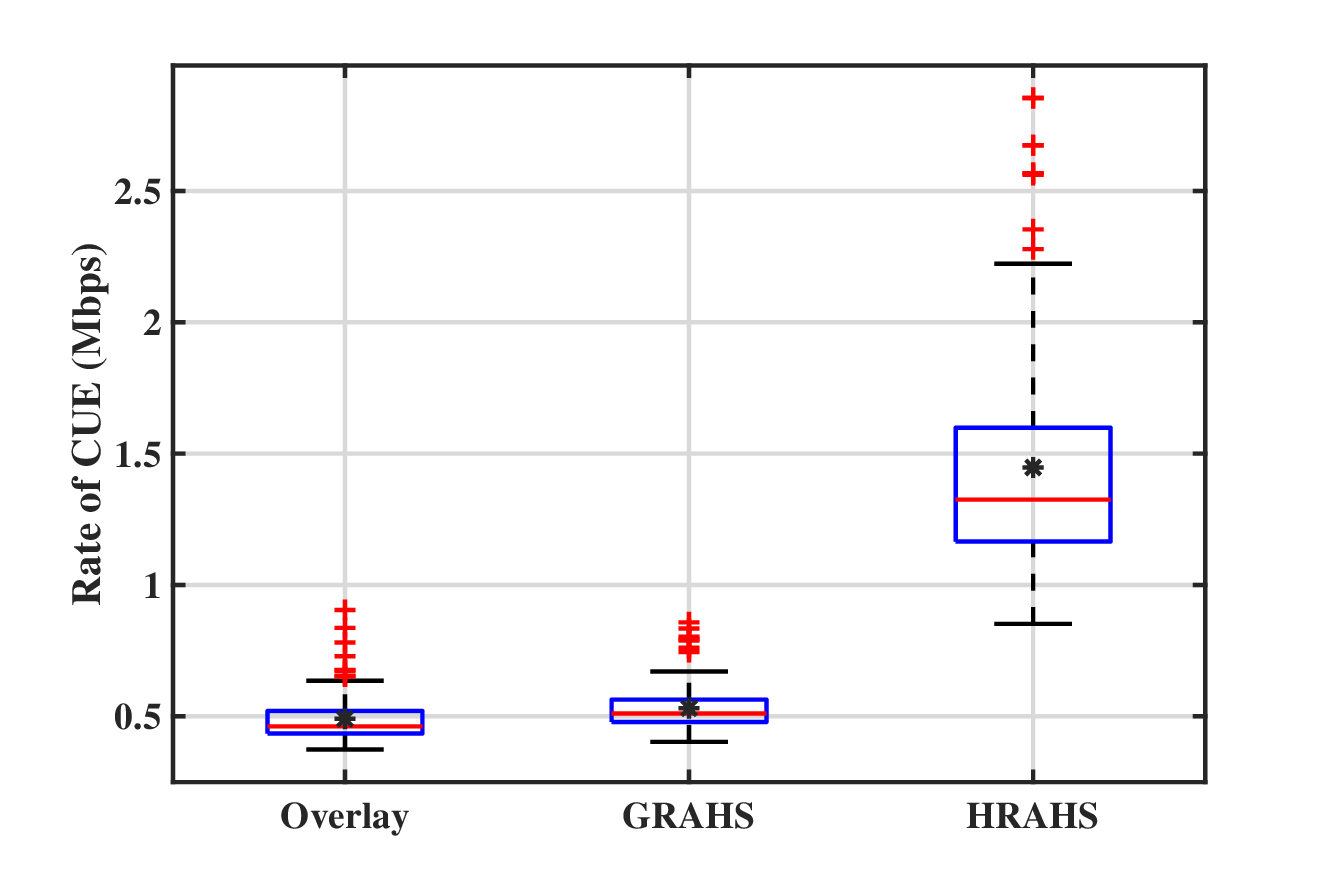}
         \caption{Sum-Rate for 126 \acp{CUE} in the system}
         \label{fig:CUE_rate}
     \end{subfigure}
        \caption{Performance of the different Resource allocation algorithms in terms of Packet Loss Rate, Average Delay, Sum-Rate experienced by the \acp{CUE}}
        \label{fig:CUEPerfeval}
\end{figure*}
\begin{figure*}[t]
     \centering
     \hspace{-0.2cm}
          \begin{subfigure}{0.31\textwidth}
         \centering
        \includegraphics[width = 0.9\textwidth]{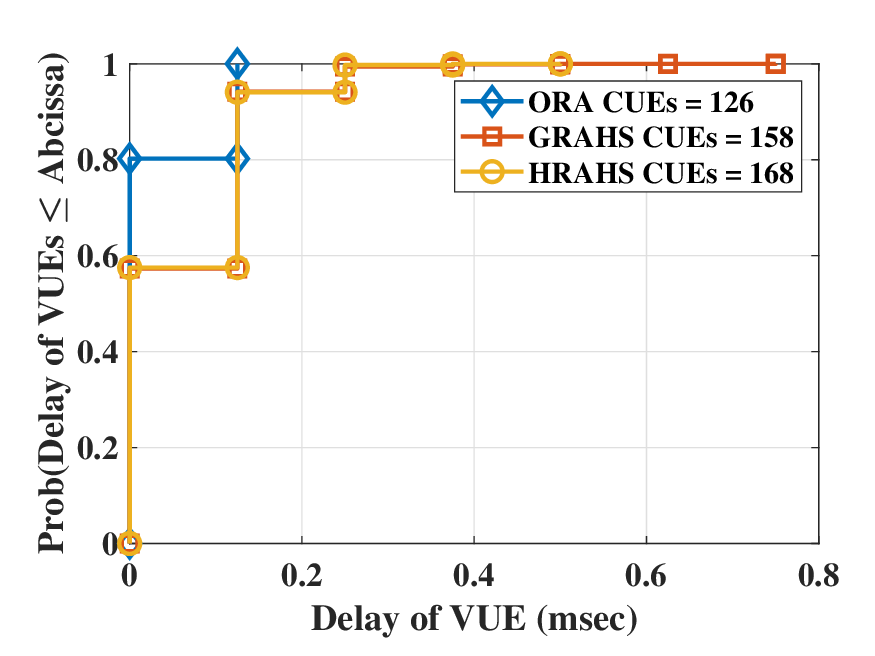} 
\caption{CDF of Avg. Delay for  VUEs }
\label{fig:VUE_delay}
     \end{subfigure}
     \hspace*{-0.4cm}
     \begin{subfigure}{0.31\textwidth}
         \centering
         \includegraphics[width = 0.9\textwidth]{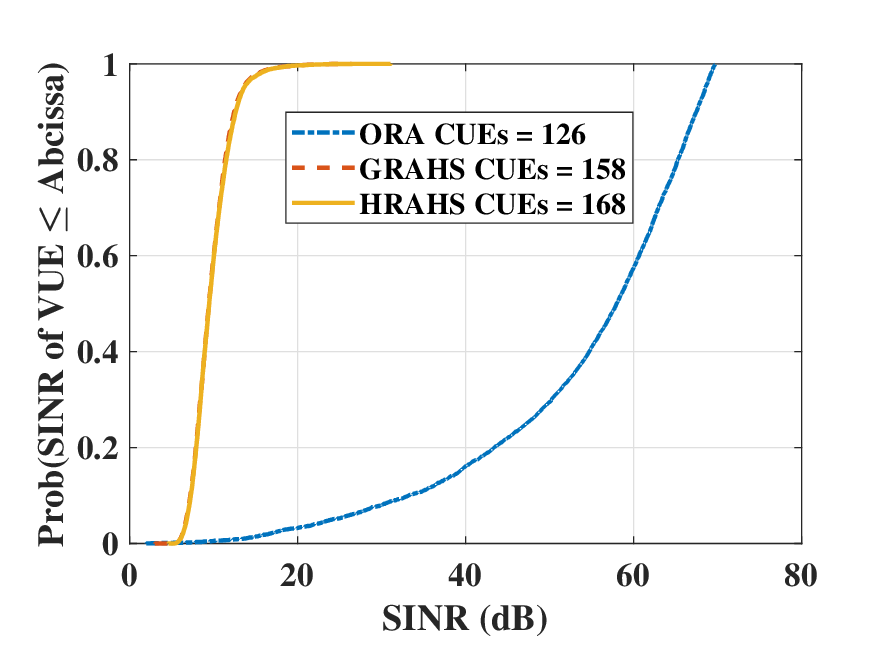}
\caption{Outage Probability of \acp{VUE}}
\label{fig:SINRVUE}
     \end{subfigure}
 \hspace*{-0.4cm}       
     \begin{subfigure}{0.31\textwidth}
         \centering
    \includegraphics[width = 0.9\textwidth]{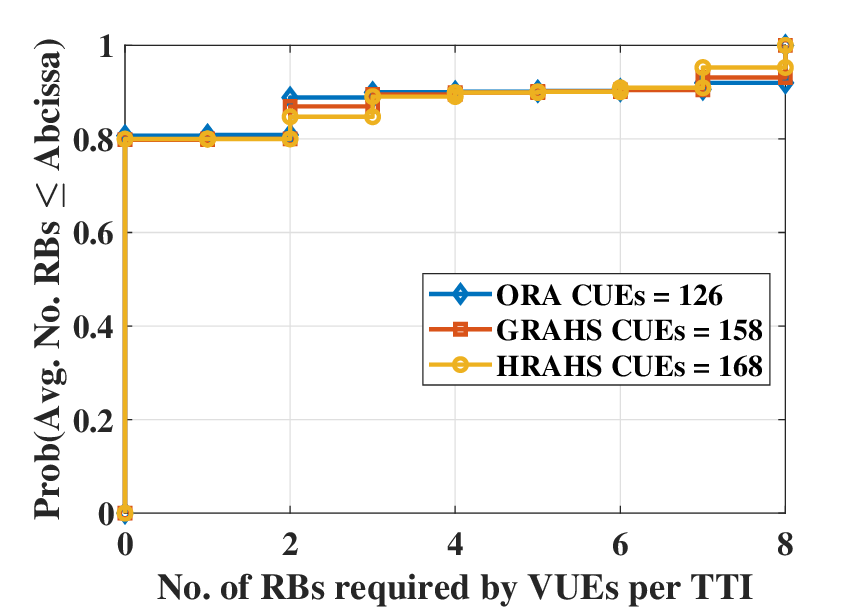}
\caption{CDF of RBs reqd. by VUEs per slot}
\label{fig:RBVUE}
     \end{subfigure}
        \centering\caption{Performance of the different Resource allocation algorithms in terms of the Average Delay, Outage Probability, and the \ac{RB} requirement of \acp{VUE}}
        \label{fig:VUEPerfeval}
\end{figure*}
To establish the need for link adaptation in resource allocation, we have used the algorithm in~\cite{Le2017}, which uses the same power allocation and resource sharing as our work, but \textit{assigns one \ac{RB} per \ac{CUE}}. We refer to this baseline algorithm from~\cite{Le2017} as \textit{``Single \ac{RB} Allocation with Hungarian Sharing" - SRAHS}. While implementing SRAHS, we schedule a \ac{CUE} when it needs a single \ac{RB}. In other words, if the \ac{CUE} has a poor channel condition and requires more than one \ac{RB}, SRAHS waits till the channel condition of the \ac{CUE} improves, and the user requires a single \ac{RB} only. This ensures that each \ac{CUE} would need one \ac{RB} only, as mentioned in \cite{Le2017}.  Fig. \ref{fig:LAutility_PLR} and Fig. \ref{fig:LAutility_delay} shows the packet loss ratio and average packet delay of SRAHS, ORA, GRAHS, HRAHS, and ORA. It is observed that with SRAHS the packet loss rate of the \acp{CUE} is more than 2\% when there are 25 \acp{CUE} in the system. In contrast, when link adaptation is used, the packet loss rate is around 2\% even when there are more users in the system - 126 \acp{CUE} for ORA, 158 \acp{CUE} for GRAHS, and 168 \acp{CUE} for HRAHS. Although the average packet delay of SRAHS for successfully delivered packets is less than the delay bound, it remains higher than link adapted resource allocation, implying that numerous packets have been dropped due to delay bound violation. Having established the need for link adaptation, we carry out the rest of the analysis using ORA, GRAHS, and HRAHS.  
\subsubsection{\textbf{Performance of the \acp{CUE}}}  

\begin{figure}[h]
\centering
\begin{subfigure}{0.25\textwidth}
         \centering
         \includegraphics[width=\textwidth]{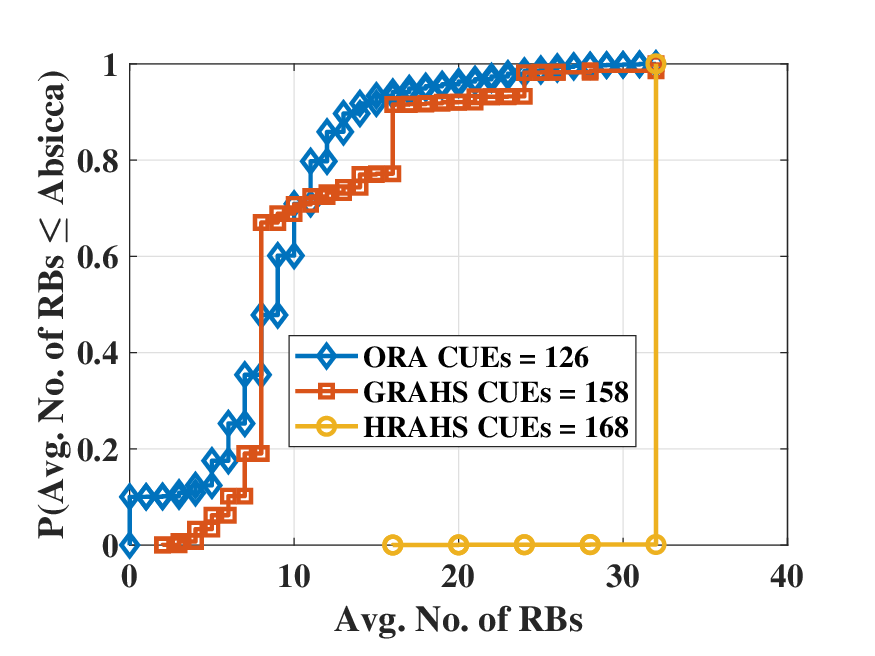}
         \caption{Occupied by all \acp{CUE}}
         \label{fig:RBCUE}
     \end{subfigure}
 \hspace*{-0.5cm}       
     \begin{subfigure}{0.25\textwidth}
         \centering
         \includegraphics[width = \textwidth]{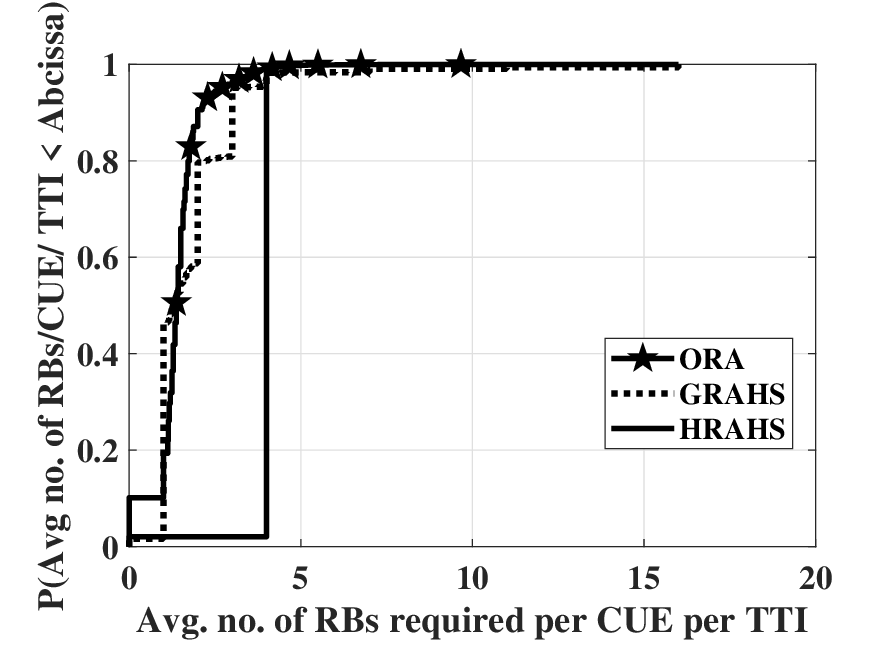}
         \caption{Occupied by each \ac{CUE}}
         \label{fig:RBperCUEperslot}
     \end{subfigure}
        \caption{CDF of the Number of \acp{RB} occupied by the \acp{CUE}. The CDF has been shown for the number of \acp{RB} occupied by all \acp{CUE} as well as per \acp{CUE}.}
        \label{fig:RB_Alloc}
\end{figure}

\indent In this section, we discuss the performance of the algorithms. Fig. \ref{fig:CUE_PLR} shows the variation of packet loss ratio with the number of \acp{CUE}, while Fig. \ref{fig:CUE_delay} shows the \ac{CDF} of the average delay experienced by the users in milliseconds. It may be observed from Fig. \ref{fig:CUE_PLR}  and Fig. \ref{fig:CUE_delay} that the number of \acp{CUE} which meet the 2\% packet loss ratio while maintaining the delay bound is 126, 158, and 168 for ORA, GRAHS, and HRAHS, respectively.   So, these values are the real-time traffic capacity of the network for the given simulation scenario. ORA supports a lower real-time traffic capacity due to the dedicated resource assignment for which the limit on the maximum number of users scheduled in a \ac{TTI} is applicable to both the \acp{CUE} and \acp{VUE}. The supported capacity increases with resource sharing between \acp{CUE} and \acp{VUE}. Furthermore, the capacity of HRAHS is even higher than GRAHS. GRAHS and HRAHS allocate resources to users in the order they appear in the TTL-sorted buffers. The following two conditions may arise with GRAHS: - 1)
If users with poor channel conditions are ahead in the queue, they will occupy a larger number of \acp{RB} thereby throttling resource availability to the trailing users. 2) If users with poor channel conditions appear later in the queue, then a sufficient number of resources may not be available. As   GRAHS aims to maximize the sum rate, it implicitly prioritizes users with better channel conditions, \textit{i.e.}, fewer resource requirements. This can be corroborated from Fig. \ref{fig:RBperCUEperslot}, which plots the average number of \acp{RB} required by each \ac{CUE} per \ac{TTI}. It is seen from Figs. \ref{fig:RBCUE} and \ref{fig:RBperCUEperslot} that GRAHS occupies fewer resources. In fact, the percentage of users requiring 1/2/3/4 \acp{RB} with GRAHS varies as 59\%/14\%/15\%/4.78\%, respectively. HRAHS, on the other hand, universally assigns four \acp{RB} per user, thereby supporting users even with average channel conditions. As a result, the real-time traffic capacity of HRAHS is 168 \acp{CUE}, which is even higher than that of GRAHS at 158 \acp{CUE}.\\
\indent Fig. \ref{fig:CUE_delay} shows  the comparison of the delay performance of the three algorithms at two levels - (i) for the same number of users, and (ii) for the real-time traffic capacity of individual algorithms. For comparing the performance with the same number of users, we have used the real-time traffic capacity of ORA of 126 users, which is the least among all the algorithms. It is observed that with the same number of users, ORA delivers the worst delay performance among all algorithms. The reason can also be attributed to the fact that for ORA the limit on the maximum number of users that can be scheduled in a \ac{TTI} applies to both \acp{CUE} and \acp{VUE}.  When the delay performance is compared at the capacity of individual algorithms, it is seen that the delay experienced with GRAHS and HRAHS increases with the number of \acp{CUE}, but it is well within the maximum allowable delay of 50ms. \\
\indent Fig. \ref{fig:CUE_rate} compares the sum-rate achieved by the three different algorithms for 126 \acp{CUE}, \textit{i.e.}, for the real-time traffic capacity of the ORA algorithm. It is seen that HRAHS is able to provide a much higher sum-rate for the \acp{CUE}, than GRAHS. HRAHS achieves the higher system capacity and improved sum-rates by efficient utilization of the radio resources. The resource utilization of the three algorithms can be observed from Fig. \ref{fig:RBCUE}, which shows the number of \acp{RB} occupied by the \acp{CUE}. It shows that HRAHS has a better resource utilization, which gets reflected in the improved capacity and sum rates.
\subsubsection{\textbf{Performance of the \acp{VUE}}} Fig. \ref{fig:VUEPerfeval} shows the performance of the \acp{VUE} under the different resource allocation schemes. In ORA, both \acp{CUE} and \acp{VUE} are sorted in the same buffer according to their time-to-live. As \ac{VUE} packets have a shorter TTL, hence when generated in the same \ac{TTI}, \ac{VUE} packets will have a higher priority than \acp{CUE}. As the ORA  algorithm implicitly prioritizes \acp{VUE} over \acp{CUE}, hence it is observed from Fig. \ref{fig:VUE_delay} that the \acp{VUE} will experience least delay when being scheduled with ORA. However, it comes at the cost of increased outage probability as seen in Fig. \ref{fig:SINRVUE}. This is because in \acp{TTI} where the \ac{CUE} packets have a higher priority, they will occupy a larger number of \acp{RB} due to their larger packet size. This will throttle the availability of bandwidth to the trailing \acp{VUE}. In addition, if these \acp{VUE} have a poorer channel condition then the number of \acp{RB} needed to serve the packet may not be available, resulting in the higher outage probability. On the other hand, when the \acp{RB} are shared, as in GRAHS and HRAHS, the \acp{VUE} have access to a larger number of \acp{RB}, resulting in the improved outage probability performance. Fig. \ref{fig:RBVUE} shows  the resource utilization of the \acp{VUE}. It remains fairly constant for all the algorithms as the number of \acp{VUE} remain constant.

\subsubsection{\textbf{Performance of the \acp{BUE}}} 
\begin{figure}[t]
\centering
\includegraphics[trim={0cm 0cm 0cm 1cm}, width=0.3\textwidth]{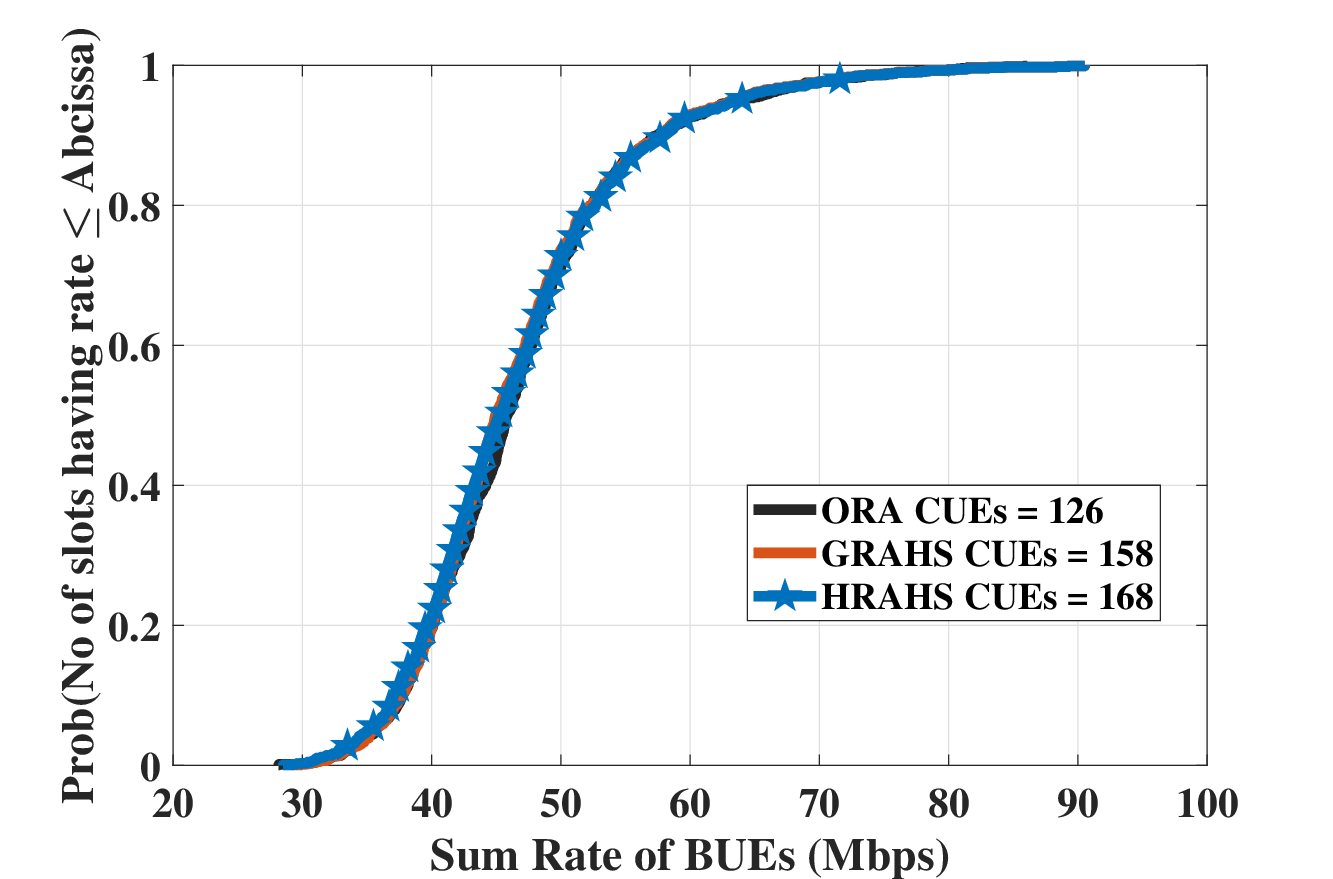}
\caption{CDF of the rate of the best effort users }
\label{fig:rateBUE}
\end{figure}
Fig. \ref{fig:rateBUE} shows CDF of the sum-rate of the \acp{BUE}. For all the three algorithms, the \acp{BUE} have been scheduled using MAX C/I algorithm in which  each \ac{RB} is assigned to that \ac{CUE} which supports the highest rate. As the  bandwidth, resource allocation, and the number of \acp{BUE} remain unchanged, hence, the \ac{CDF} performance remains the same for all the three cases. The existence of the \acp{BUE} necessitates the use of multiple numerologies.\\
\indent Therefore, we may infer that link adaptation is  indispensable for serving a combined subscriber base of \ac{V2I} and \ac{V2V} users in a \ac{C-V2X} system operating with 5G technology.

\section{Conclusion}\label{sec:Conclusion}
\acresetall
In this work, we have asserted the need for link adaptation in spectrum allocation of \ac{C-V2X} systems. This work aims to increase the number of \ac{V2I} users for a given number of \ac{V2V}, and best-effort users. The diverse \ac{QoS} requirements of these users have been provided using the multi-numerology-based frame structure of 5G. The \ac{V2V} users operate in underlaid mode, sharing resources with the \ac{V2I} users. In contrast to existing works,
 our method assumes that the number of \acp{RB} required by the users is a function of their underlying channel conditions and \ac{QoS} requirements.  Our first algorithm, called GRAHS, assigns resources to users using a greedy method, and the second one assigns resources to \ac{V2I} users using the Hungarian method. In their second stage, both algorithms use the Hungarian algorithm for optimal pairing between \ac{V2V} and \ac{V2I} users.  We have observed that link adaptation significantly improves the QoS provisioning, and, hence, the supported real-time traffic capacity of the systems. \\
 \indent In this work, we have considered all users to have a constant velocity and have aimed to maximize the \ac{V2I} links only. As a future extension, we plan to design lower computationally complex resource allocation algorithms, which in addition to increasing the number of \ac{V2I} users will also aim to increase the number of \ac{V2V} links for different mobility conditions. Furthermore, these resource allocation algorithms will also aim to use fewer resources for the QoS-constrained users to improve the sum rate of the best-effort users. 

\printbibliography



\begin{acronym}
	\acro{2G}{2$^\text{nd}$ Generation}
	\acro{3G}{3$^\text{rd}$ Generation}
	\acro{4G}{4$^\text{th}$ Generation}
	\acro{5G}{5$^\text{th}$ Generation}
	\acro{3GPP}{$\text{3}^\text{rd}$ Generation Partnership Project}
	\acro{A3C}{Actor-Critic}
	\acro{ABR}{Adaptive Bitrate}
	\acro{BS}{Base Station}
	\acro{BUE}{Best-Effort UE}
	\acro{BWP}{Bandwidth Partition}
	\acro{CDN}{Content Distribution Network}
	\acro{CDF}{Cumulative Distribution Function}
	\acro{CSI}{Channel State Information}
	\acro{CUE}{Cellular User Equipment}
	\acro{DASH}{Dynamic Adaptive Streaming over HTTP}
	\acro{DL}{deep learning}
 \acro{DRL}{Deep Reinforcement Learning}
	\acro{DRX}{Discontinuous Reception}
	\acro{D2D}{Device-to-Device}
	\acro{EDGE}{Enhanced Data Rates for \ac{GSM} Evolution.}
			\acro{gNB}{general NodeB}
	\acro{eNB}{evolved NodeB}
	\acro{GSM}{Global System for Mobile}
	\acro{FL}{Federated Learning}
	\acro{TL}{Transfer Learning}
	\acro{HD}{High Definition}
	\acro{HSPA}{High Speed Packet Access}
	\acro{LSTM}{Long Short Term Memory}
	\acro{LTE}{Long Term Evolution}
	\acro{ML}{machine Learning}
	\acro{MTL}{Multi-Task Learning}
	\acro{MCS}{Modulation and Coding Scheme}
	\acro{NSA}{Non-Standalone}
	\acro{HVPM}{High voltage Power Monitor}
 \acro{PUE}{Pedestrian UE}
	\acro{QoS}{Quality of Service}
	\acro{QoE}{Quality of Experience}
	\acro{RB}{Resource Block}
	\acro{RF}{Random Forest}
	\acro{RFL}{Random Forest}
	\acro{RL}{Reinforcement Learning}
	\acro{RRC}{Radio Resource Control}
	\acro{RSSI}{Received Signal Strength Indicator}
	\acro{RSRP}{Reference Signal Received Power}
	\acro{RSRQ}{Reference Signal Received Quality}
	\acro{SCS}{Sub-carrier \ Spacing}
	\acro{SINR}{Signal-to-Interference-Plus-Noise-Ratio}
	\acro{SNR}{Signal-to-Noise-Ratio}
	\acro{SE}{spectral efficiency}
	\acro{UE}{User Equipment}
	\acro{UHD}{Ultra HD}
	\acro{VUE}{Vehicular UE}
	\acro{VoLTE}{Voice over LTE}
	\acro{RNN}{Recurrent Neural Network}
	\acro{WiFi}{Wireless Fidelity}
	\acro{ARIMA}{Auto Regressive Integrated Moving Average}
	\acro{ML}{Machine Learning}
	\acro{NR}{New Radio}
	\acro{Non-IID} {Independent and Identically Distributed}
 \acro{OFDMA}{Orthogonal Frequency Division Multiple Access}
	\acro{TP}{Throughput Prediction}
	\acro{CQI}{Channel Quality Indicator}
	\acro{URLLC}{Ultra Reliable Low Latency Communications}
	\acro{V2X}{Vehicle-to-Everything}
	\acro{C-V2X}{Cellular Vehicle-to-Everything}
	\acro{V2I}{Vehicle-to-Infrastructure}
	\acro{V2V}{Vehicle-to-Vehicle}
    \acro{OFDM}{Orthogonal Frequency Division Multiplexing }
    \acro{BWP}{Bandwidth Part}
    \acro{PUCCH}{Physical Uplink Control Channel}
    \acro{TTL}{Time-to-Live}
    \acro{RA}{resource allocation}
    \acro{BLER}{Block Error Rate}
    \acro{LA}{Link Adaptation}
    \acro{RC}{Resource Chunk}
    \acro{TTI}{Transmission Time Interval}
\end{acronym}
\end{document}